\documentclass[
  journal=pasa,
  manuscript=research-paper, 
  year=2024,
  volume=,
]{cup-journal}

\usepackage{graphicx}
\usepackage{xcolor}
\usepackage{microtype,siunitx}
\usepackage{amssymb}
\usepackage{amsmath}
\usepackage{xfrac}
\usepackage[backref=page]{hyperref}
\usepackage{multirow}
\definecolor{dodgerblue}{RGB}{30, 144, 255}
\definecolor{crimson}{RGB}{220, 20, 60}
\definecolor{darkerblue}{RGB}{0, 0, 139}
\hypersetup{colorlinks,citecolor=dodgerblue,linkcolor=blue,urlcolor=darkerblue}
\usepackage{enumitem, xcolor}
\let\svitem\item

\usepackage{booktabs} 
\usepackage{threeparttable,longtable}  
\usepackage{makecell}
\usepackage{tabularx}
\usepackage{array}

\usepackage[pagewise]{lineno}

\usepackage[skip=0.5ex]{subcaption}
\usepackage{cprotect}
\title{X-ray and CO-Derived Column Densities in AGN: A Study of Obscuration Properties in CTAGN and Non-CTAGN}

\author{M. L. H. Musa}
\affiliation{Department of Physics, Faculty of Science, Universiti Malaya, 50603 Kuala Lumpur, Federal Territory of Kuala Lumpur, Malaysia}

\author{Z. Z. Abidin}
\affiliation{Department of Physics, Faculty of Science, Universiti Malaya, 50603 Kuala Lumpur, Federal Territory of Kuala Lumpur, Malaysia}
\email[Zamri Zainal Abidin]{zzaa@um.edu.my}

\author{A. Annuar}
\affiliation{Department of Applied Physics, Faculty of Science and Technology, Universiti Kebangsaan Malaysia, 43600 UKM Bangi, Selangor, Malaysia}

\author{D. A. A. Lee}
\affiliation{Department of Physics, Faculty of Science, Universiti Malaya, 50603 Kuala Lumpur, Federal Territory of Kuala Lumpur, Malaysia}

\doi{10.1017/pasa.20XX.XX}

\received {DD Mmm YYYY}
\revised  {DD Mmm YYYY}
\accepted {DD Mmm YYYY}
\published{DD Mmm YYYY}


\begin{document}

\sloppy
\begin{abstract}
Obscuration in active galactic nuclei (AGN) provides valuable insights into the nature of the material surrounding the central engine. Compton-thick AGN (CTAGN), characterised by a column density of $N_{\mathrm{H}} \geq 1.5 \times 10^{24} \ \mathrm{cm}^{-2}$, are heavily obscured by substantial amounts of dust and gas. While X-ray observations are primarily used to determine this column density, our understanding of obscuration properties in the sub-mm regime, particularly for CTAGN, remains limited. In this study, we analyse archival data from the Atacama Large Millimetre/sub-millimetre Array (ALMA) for both CTAGN and non-CTAGN sources, as identified by the 70-month catalogue of the all-sky hard X-ray \textit{Swift}/Burst Alert Monitor survey and other X-ray surveys. Integrated intensity maps (moment 0) of CO(3--2) emission reveal a concentrated distribution of dense gas around the nucleus. Utilising a constant CO-to-$\mathrm{H_2}$ conversion factor, $X_{\mathrm{CO}} = 2.2 \times 10^{20} \ \mathrm{cm}^{-2} \ (\mathrm{K\ km\ s}^{-1})^{-1}$, we find that the derived molecular hydrogen column densities, $N_{\mathrm{H_2}}$, are generally lower than the total hydrogen column densities, $N_{\mathrm{H}}$, obtained from X-ray observations. However, the $N_{\mathrm{H_2}}$ values derived in this work are slightly higher than those reported in previous studies due to the adoption of a higher CO-to-$\mathrm{H_2}$ conversion factor. This discrepancy between $N_{\mathrm{H}}$ and $N_{\mathrm{H_2}}$ is consistent with prior findings that X-ray-derived column densities are typically higher, except in the case of non-CTAGN, where $N_{\mathrm{H_2}}$ can exceed $N_{\mathrm{H}}$. Statistical analysis using Kendall and Spearman tests reveals a positive monotonic relationship between $N_{\mathrm{H}}$ and $N_{\mathrm{H_2}}$, although the correlation is not statistically significant. This suggests a complex interplay of factors influencing these properties. The optically thick nature of CO in dense regions may contribute to the observed discrepancies. Our results highlight the importance of adopting an accurate CO-to-$\mathrm{H_2}$ conversion factor to derive reliable column densities, which could serve as an alternative method for identifying CTAGN. Further investigations with more comprehensive data sets and refined methodologies are needed to better understand the relationship between sub-millimetre and X-ray properties in AGNs.
\end{abstract}

\section{Introduction}
\label{sec:intro}

Active galactic nuclei (AGN) are extremely bright compact regions at the centres 3\% to 10\% of galaxies \citep{Gatica2023} and are more luminous ($L_{\rm{bol}} \approx 10^{48} \ \rm{erg \ s^{-1}}$) than normal galaxies \citep{Fabian1999}. It is believed to play such a significant role in galaxy evolution \citep{Koemendy1995, Kormendy2013}. Supermassive black holes (SMBHs) with masses, $M_{\rm{SMBH}} \ \gtrsim 10^6 \ M_{\odot}$ are thought to power these regions. It is assumed that the intense radiation emitted by AGN is caused by material falling into the black hole, which heats up and emits energy as it is pulled in. AGN have been linked to the establishment of scaling relations between massive black holes (MBHs) and the properties of their host galaxies. They also regulate cooling flows in galaxy clusters, drive outflows across multiple scales, from the accretion disk to the circumgalactic medium, and contribute to the development of the red sequence in massive galaxies \citep{kormendy2013coevolution,heckman2014coevolution}. The luminosities range from about $10^{40}$ erg/s to $10^{47}$ erg/s for distant quasars. One key aspect of AGN is obscuration, which refers to the amount of dust and gas that blocks our view of the nucleus. The obscuration can be characterised by the column density, $N_{\rm{H}}$ of the materials that lie along the line of sight.

 Compton-thick AGN (hereafter, CTAGN) is a specific type of AGN where the nuclear emission is obscured by a significant amount of dust and gas. The intense emissions from AGNs, including X-rays, originate primarily from energetic particles surrounding the hot corona, which are accelerated and confined by magnetic fields at the accretion disk \citep{merloni2001}. These X-rays are often Compton-scattered to longer wavelengths \citep{Antonucci1993-ge}, as they interact with dense clouds of dust and gas. In extreme cases, these X-ray emissions can be completely hidden from our observations because of heavy obscuration by dust and gas. Type 2 AGN, which include Seyfert 2 galaxies and type 2 quasars, are characterised by significant obscuration of the central engine. This obscuration is often attributed to their edge-on orientation relative to our line of sight, consistent with the AGN unification model \citep{Antonucci1993-ge}. Understanding these sources can provide insights into the composition and dynamics of material surrounding the central engine, as well as how this material influences AGN obscuration.

Traditionally, identifying CTAGNs has been primarily accomplished through X-ray observations using instruments such as Chandra, XMM-Newton, and NuSTAR, where the column densities were derived from X-ray absorption. Since X-ray observations are reliable against point source objects, for example, AGNs, we can easily distinguish them from other objects. However, sometimes other powerful objects, such as X-ray binaries, could also have the potential to contaminate these sources, despite also being often limited by the sensitivity of the instruments. For sources to be classified as AGN, the flux of X-ray point sources must be higher than $10^{42} \rm{erg \ s^{-1}}$, which is not the case for starburst galaxies (< $10^{42} \ \rm{erg \ s^{-1}}$) \citep{Jackson2010-da}. For a CTAGN, additional parameters need to be considered as well, which are the column densities, which it must be equal to or larger than the inverse of the Thomson cross-section $N_{\mathrm{H}} \geq 1.5 \times 10^{24} \ \rm{cm^{-2}}$ \citep{Comastri2004}. The relationship between X-ray absorption and hydrogen column density is characterised by the fact that the hydrogen column densities, $N_{\rm{H_2}}$ inferred from X-ray absorption are typically 5-30 times higher than the neutral atomic hydrogen column densities, $N_{\rm{HI}}$ derived from $\lambda$21 cm HI absorption toward radio-loud AGN \citep{Liszt2021}.

The relationship between X-ray-derived hydrogen column density, $N_{\rm{_H}}$ and HI column density, $N_{\rm{HI}}$ is significant for understanding obscuration in AGN. Studies have shown that X-ray-derived column densities can often exceed total HI column densities along the same line of sight, indicating that additional sources of obscuration, such as molecular hydrogen or ionised gas, may be present \citep{Liszt2021}. This discrepancy emphasises the need for a comprehensive understanding of all gas components in AGN environments. Furthermore, a positive correlation between X-ray absorption and HI column density has been observed, suggesting that, as the X-ray column density increases, so does the HI column density, although this relationship may not always be statistically significant.

In recent years, there has been growing interest in finding alternative methods to measure the column density of AGN \citetext{e.g. \citealp{Silver_2023,Silverman_2023, Torres_Alb__2023}}. One such approach is to use molecular hydrogen, $\rm{H_2}$ which can provide a more accurate representation of the obscuration properties. It is important to note that there are different types of molecular gas being discussed. Warm $\rm{H_2}$, typically observed in the near-infrared and associated with temperatures ranging from 500 to 2000 K, differs significantly from cold molecular gas, which is often inferred from CO observations and generally exists at much lower temperatures (approximately 20 to 200 K).

Warm $\rm{H_2}$ is typically found in regions of active star formation, where it can participate in processes that lead to the formation of stars. The higher temperatures of warm $\rm{H_2}$ allow it to be excited by nearby stellar radiation, making it a key player in the dynamics of star-forming regions. In contrast, cold $\rm{H_2}$ serves as the primary reservoir for star formation and is often found in dense molecular clouds, where conditions are favourable for gravitational collapse. CO does not always trace the same regions as warm or hot $\rm{H_2}$, making it essential to consider these differences when interpreting the physical conditions in AGN environments. While CO is a reliable tracer for cold, dense molecular gas, it primarily indicates regions that are cooler and may not capture the dynamics occurring in warmer gas phases. This distinction is crucial because CO emission can sometimes arise from different physical conditions than those represented by warm $\rm{H_2}$.

Although measuring the molecular hydrogen, $\rm{H_2}$ is quite challenging due to its weak emission, tracers such as CO can be used. CO is particularly useful for tracing the cold, dense molecular gas that forms stars and extreme regions such as AGN \citetext{e.g. \citealp{davies2012dense, esposito2022agn, esposito2024modelling}}. Due to its abundance and easy observability, CO also can provide crucial information about the physical properties of molecular clouds, such as their mass, density, temperature, and velocity structure. \citet{Hicks2009-nx} discussed the role of molecular hydrogen in dense environments, particularly in obscuring regions in the central region of AGN. Their work highlights that the molecular hydrogen, $\rm{H_2}$ can exceed the value ranges from $10^{22}$ up to $10^{24} \rm{cm^{-2}}$, with the inner radius of 100 pc in AGN. 

In this work, we present a preliminary approach to explore the gap between total hydrogen column density, $N_{\rm{HTotal}}$ as derived from X-ray observation and molecular hydrogen column density, $N_{\rm{{H_2}}}$, as described from sub-millimetre observation. We aim to assess whether molecular hydrogen column density $N_{\rm{H_2}}$ could serve as an alternative indicator of obscuration in AGN, and particularly help to search for CTAGN. Understanding this relationship could improve the method of identifying heavily obscured AGNs, as well as their nature of obscuration. 

The paper is organised as follows: We describe the Atacama Large Millimetre/sub-millimetre Array (ALMA) observations of CTAGNs as well non-CTAGNs in section~\ref{sec: Observations and Data Reduction}. In the same section, we discuss the steps in reducing the data obtained from the archive. In section~\ref{sec: result}, we present our main results, discussing the column density derived, and the correlation with X-ray column density. Finally, our conclusions are summarised in section~\ref{sec: conclusion}. In this paper, we adopt a $\Lambda$CDM cosmology with parameters $H_0 = 70 \, \text{km} \, \text{s}^{-1} \, \text{Mpc}^{-1}$, $\Omega_m = 0.3$, and $\Omega_\Lambda = 0.7$.

\section{Observations and Data Reduction}
\label{sec: Observations and Data Reduction}
\subsection{AGN Samples}
In this study, we examine a sample of AGN characterised by a redshift, $z \leq 0.005$, while also distinguishing between CTAGN and non-CTAGN. This allows us to achieve high spatial resolution for imaging and spectroscopic data in the central regions of galaxies, where the dust that causes obscuration is present. At low redshift, spectral line measurements such as CO are less affected by redshift; thus it will improve the signal-to-noise ratio. Further considerations are their morphology (all are spirals, oriented roughly face-on towards the observer). The AGN sample was derived from the 70-month catalogue of the all-sky hard X-ray \textit{Swift}/Burst Alert Monitor (\textit{Swift}/BAT) survey \citep{baumgartner201370}, as reported by \citet{Ricci2017}, which identified 838 AGNs, including 55 CTAGNs from \citet{Ricci2015}. Subsequently, we searched for corresponding ALMA observations in the ALMA Science Archive, specifically targeting data on the CO(3-2) transition. Our investigation revealed a limited availability of data, with only five datasets corresponding to CTAGN and two datasets for non-CTAGN. To enhance comparability, we identified an additional three non-CTAGN sources from the Chandra ACIS Survey for X-Ray AGN in Nearby Galaxies (CHANSNGCAT, \citealp{she2017chandra}). As a result, we have 12 AGNs in total with 6 samples each for both CTAGN and non-CTAGN.

We utilised the NASA/IPAC Extragalactic Database (NED) to gather various properties of galaxies, including their celestial coordinates (RA and Dec), redshift values, morphology, and the types of AGN they contain. To ensure consistency, we crosschecked the AGN types obtained from NED/IPAC with data from The 70-Month \textit{SWIFT}/BAT catalogue. Additionally, we sourced other properties, such as X-ray column density ($N_{\rm{H}}$) and AGN location from past literature. Table \ref{tab:properties} outlines the sample properties analysed in this study for easy reference

\begin{table*}[hbt!]
\caption{Properties of the AGN sample used in this work, were obtained from the NASA/IPAC Extragalactic Database (NED).}
\label{tab:properties}
\begin{threeparttable}
    \begin{tabular}{lccccccc}
    \toprule
    Source Name & RA$\rm{_{J2000}} $ & Dec$\rm{_{J2000}}$  & Morphology  & AGN Type & $z$ \\ 
    & (h:m:s) & (d:m:s) & & & \\
    \midrule
    \textbf{Compton-Thick AGN} & \\
    Circinus Galaxy & 14:13:09.950 & -65:20:21.20 & SA(s)b: & Sy2 & 0.00140 \\
    NGC 1068 & 02:42:40.711 & -00:00:47.81 & (R)SA(rs)b & Sy2 & 0.00380  \\ 
    NGC 4945 & 13:05:27.477 & -49:28:05.57 & SB(s)cd: edge-on & Sy2 & 0.00190  \\ 
    NGC 5643 & 14:32:40.743 & -44:10:27.86 & SAB(rs)c & Sy2 & 0.00399  \\ 
    NGC 6240 & 16:52:58.871 & +02:24:03.33 & I0: pec & Sy1.9 & 0.00245 \\ 
    NGC 7582 & 23:18:23.500 & -42:22:14.00 & (R')SB(s)ab & Sy2 & 0.00525 \\ 
    \midrule
    \textbf{Non-Compton-Thick AGN} \\							
    NGC 613 & 01:34:18.170 & -29:25:06.10 & SB(rs)bc & Sy/LINER & 0.00500 \\ 
    NGC 1097 & 02:46:19.05 & -30:16:29.6 & SB(s)b & LINER b & 0.00424 \\ 
    NGC 1566 & 04:20:00.42 & -54:56:16.1 & SAB(s)bc & Sy1.5 & 0.00500 \\ 
    NGC 1808 & 05:07:42.34 & -37:30:47.0 & (R)SAB(s)a & - & 0.00334 \\ 
    NGC 6300 & 17:16:59.47 & -62:49:14.0 & SB(rs)b & Sy2 & 0.00370 \\ 
    NGC 7314 & 22:35:46.19 & -26:03:01.68 & SAB(rs)bc & Sy1.9 & 0.00480 \\ 
    \bottomrule
    \end{tabular}
\end{threeparttable}
\end{table*}


\subsection{ALMA Observations}
We utilised ALMA Band 7 (275 - 373 GHz) to observe the CO(3-2) transition at a rest frequency of 345.8 GHz in the central regions of our AGN sample. CO(3-2) is particularly effective for tracing dense gas in AGN environments due to its higher critical density ($n_{crit} = 3.6 \times 10^4 \ \rm{cm^{-3}}$; \citealp{carilli2013cool}) compared to other CO transitions, such as CO(1-0) and CO(2-1). This high density sensitivity is important because Compton-thick AGNs (CTAGNs) are typically obscured by thick layers of dust and gas. Additionally, the CO(3-2) transition has an excitation temperature of around 33 K (\citealp{Wilson2009}), making it suitable for probing the relatively warm, dense molecular gas that characterises the obscuring regions in CTAGNs. Our analysis primarily relied on archival data sourced from the ALMA Science Archive. We systematically queried and acquired the pipeline data, thereby obtaining the cube file necessary for subsequent in-depth analysis. Table \ref{tab:alma_observation_details} lists all the ALMA observation properties obtained from the ALMA Science Archive.

\newcolumntype{C}[1]{>{\raggedright\centering\arraybackslash}p{#1}} 

\begin{table*}[htbp]
    \centering
    \caption{ALMA Observation Properties}
    \label{tab:alma_observation_details}
    \begin{tabular}{l c C{5em} C{5em} C{5em} C{4em} C{5em} C{5em} C{4em}}
    \toprule
    Source Name & Project ID & Observation Date & Angular Resolution & Velocity Resolution & FOV & Continuum Sensitivity  & Line Sensitivity & Vlsr Used \\
    & & & (arcsec) & (km/s) & (arcsec) & (mJy/beam) & (mJy/beam) & (km/s) \\
    \midrule
    \multicolumn{9}{l}{\textbf{Compton-Thick AGN}} \\
    Circinus Galaxy & 2015.1.01286.S & 2015-12-31 & 0.910 & 6.587 & 61.511 & 0.4602 & 11.282 & 433.79 \\
    NGC 1068 & 2016.1.00232.S & 2017-09-08 & 0.033 & 3.302 & 16.593 & 0.0306 & 0.686 & 1137.78 \\
    NGC 4945 & 2018.1.01236.S & 2018-12-08 & 0.317 & 1.635 & 16.659 & 0.0341 & 0.923 & 537.73 \\
    NGC 5643 & 2017.1.00082.S & 2018-04-19 & 0.462 & 3.303 & 16.690 & 0.0496 & 1.317 & 1191.43 \\
    NGC6240 & 2013.1.00813.S & 2015-05-04 & 0.176 & 6.649 & 16.907 & 0.1258 & 3.098 & 7118.11 \\
    NGC 7582 & 2017.1.00082.S & 2018-08-30 & 0.329 & 3.307 & 16.674 & 0.0687 & 1.587 & 1569.91 \\
    \midrule
    \multicolumn{9}{l}{\textbf{Non-Compton-Thick AGN}} \\
    NGC 613 & 2016.1.00296.S & 2016-11-23 & 0.294 & 0.827 & 16.705 & 0.0549 & 1.385 & 1477.01 \\
    NGC 1097 & 2015.1.00126.S & 2016-09-14 & 0.099 & 3.301 & 16.694 & 0.0304 & 0.748 & 1259.22 \\
    NGC 1566 & 2016.1.00296.S & 2016-11-24 & 0.315 & 0.827 & 16.706 & 0.0608 & 1.532 & 1502.42 \\
    NGC 1808 & 2015.1.00404.S & 2016-08-12 & 0.150 & 3.298 & 16.685 & 0.0632 & 1.538 & 998.80 \\
    NGC 6300 & 2017.1.00082.S & 2018-08-22 & 0.544 & 3.302 & 16.684 & 0.0576 & 1.393 & 1100.19 \\
    NGC 7314 & 2017.1.00082.S & 2018-08-30 & 0.330 & 3.306 & 16.703 & 0.0682 & 1.616 & 1423.51 \\
    \bottomrule
    \end{tabular}
\end{table*}

\subsubsection{Circinus Galaxy}
Circinus Galaxy is a Seyfert 2 galaxy, hosting bright active galactic nuclei. Positioned 4 degrees below the Galactic plane, this galaxy is approximately 4.5 Mpc \citep{Freeman1997}, making it one of the closest major galaxies to our own. It is obscured by ~100 pc scale dust lanes located at the galaxy plane \citetext{e.g.\citealp{Wilson2000-vw, Prieto2004-as, Mezcua2016}} and it became the ideal candidate for Compton-Thick galactic nuclei where the central emission suffers heavy obscuration and absorption along our line of sight. \citet{Greenhill2003} observed $\mathrm{H_2O}$ maser emission where this traced a warped, edge-on accretion disk close to 0.1 pc from the nucleus. They provide initial evidence of dusty and high-density molecular material around the nucleus. They also suggested that this structure may be applied to other Seyfert 2 galaxies. A similar study by \citet{Hagiwara2021-zc} investigated $\rm{H_2O}$ maser activity in the central regions of the AGN galaxies Circinus and NGC 4945 using ALMA at 321 GHz. They observed that the maser emissions in NGC 4945 show significant variability over 4-5 year timescales, with the nuclear continuum emission varying in correlation with the maser. However, this correlation is less certain in the Circinus galaxy, where the relationship between maser and nuclear continuum emissions remains unclear.

The transition of CO(3-2) was observed in the central region of the Circinus galaxy by ALMA cycle 3 project (PID: 2015.1.01286.S, PI: Franceso Costagliola). This utilised ALMA band 7 ($\sim$350 GHz) receiver with the continuum sensitivity of 0.4602 mJy/beam. This galaxy was observed on 31$\mathrm{^{st}}$ December 2015 and the pointing was centred at RA = $\mathrm{14^h 13^m 9.962^s}$, Dec = $\rm{65^d 20^m 20.937^s}$. The data were calibrated using the Common Astronomy Software Applications Package (CASA) version 4.5. This project aimed to study the dense interstellar medium of obscured AGN, at the centre of Circinus Galaxy.

\subsubsection{NGC 1068}
The galaxy NGC 1068 is one of the nearest Type 2 Seyfert galaxies to the Milky Way, with the Circinus galaxy being the closest known Seyfert. NGC 1068 is located in the constellation Cetus at a distance of approximately 14 Mpc \citep{bland1997ringberg}. \citet{Greenhill1996-wi} produced the first VLBI synthesis images of the $\mathrm{H_2O}$ water maser emission originating from the galaxy's central engine. Among prior studies, \citet{Marinucci2015} analysed NuSTAR and XMM-Newton data and reported no spectral variations below 10 keV. Furthermore, numerous investigations have focused on the nuclear kinematics and molecular content of NGC 1068 \citetext{e.g., \citealp{jaffe2004central, garcia2016alma,garcia2017alma, Garcia-Burillo2019-qo, pfuhl2020image}}, revealing distinct components: molecular gas emitting in the infrared, an X-ray-obscuring torus, and ionised gas \citep{wang2012chandra}. Using ALMA CO(3–2) observations, \citet{imanishi2018alma} reported the detection of a rotating, elongated dense molecular emission associated with the torus in the nuclear region of NGC 1068. Their findings suggest that the molecular emission rotates predominantly in the east-west direction and exhibits significant inhomogeneity, with the western region of the AGN being notably more turbulent than the eastern region.

For ALMA CO(3-2) observation, the data was taken from archival data (PID: 2016.1.00232.S, PI: Santiago Garcia-Burillo). This project aims at the circumnuclear region of NGC 1068, covering the region from $\sim$100 pc radius down to 7$\sim$ 10 pc of torus diameter. The continuum sensitivity of this target was 0.0306 mJy/beam. This observation was conducted on 8th September 2017 with the telescope pointing at RA = $\rm{2^h 42^m 40.711^s}$, Dec = $\rm{-00^d 00^m 47.840^s}$. The data were calibrated using CASA version 4.7.2.

\subsubsection{NGC 4945}
NGC 4945 is a nearby edge-on Seyfert II galaxy with a visible obscuring dust lane in the central region. It is one of the brightest radio-quiet Seyfert in 100 keV surveys \citep{Done1996-pp} and displays both Seyfert \citep{schurch2002high} and starburst \citep{bendo2016, emig2020super}.A recent study by \citet{Bolatto2021} using ALMA has detected molecular outflows of CO (3-2), $\rm{HCO^+}$ and HCN in the central region of NGC 4945. The molecular outflow plumes align with the edges of the ionised gas outflow, tracing a receding cone that is obscured behind the galaxy disk and undetectable in optical or soft X-ray emission. Another study using ALMA Band 3 observation was carried out by \citet{Henkel2018} to study the structure, dynamics, and composition of NGC 4945. Most observed molecular lines show absorption near the nuclear millimetre-wave continuum source, especially between –60 and +90 km s$^{-1}$ from the systemic velocity of 571 km s$^{-1}$. Only the CH$_3$C$_2$H J=5$\rightarrow$4 transition is unaffected, likely due to its lower critical density, with line emission absorbed slightly southwest of the continuum peak.

This galaxy is observed by ALMA (PID: 2018.1.01236.S, PI: Adam Leroy) on December 8, 2018, using a telescope pointing centred on a right ascension RA = \(\rm{13^h 05^m 27.476^s}\) and a declination Dec = \(\rm{-49^d 28^m 05.100^s}\). Data acquisition achieved a continuum sensitivity of 0.0341 mJy/beam. The calibration was performed using CASA version 5.4.0-68. The project aims to observe the nuclear starburst region of NGC 4945 to confirm and study $\sim$20 candidate super star clusters identified in previous ALMA imaging. They will measure the clusters' dynamical mass, gas mass, dust opacity, and stellar content by utilising ALMA Band 3 and 7.

\subsubsection{NGC 5643}
NGC 5643, located in the constellation Lupus, is an active galaxy that has been extensively studied, particularly in optical and X-ray wavelengths. For instance, \citet{Annuar2015} utilised NuSTAR observations, in combination with previous X-ray data, to directly measure the obscuring column and confirm the classification of the galaxy's X-ray source. \citet{garcia2021multiphase} employed ALMA CO(2–1) observations to study the cold molecular gas and MUSE IFU optical emission lines to investigate the ionised gas, aiming to explore the multiphase feedback processes at the central AGN of NGC 5643. Similarly, \citet{AlonsoHerrero2018} analysed ALMA Band 6 \(^{12}\)CO(2–1) line and rest-frame 232 GHz continuum observations, revealing that the \(^{12}\)CO(2–1) kinematics can be modelled as a rotating disk, which appears tilted relative to the galaxy's large-scale disk. Additionally, they identified strong non-circular motions in the central 0.2–0.3 arcseconds of the galaxy, with velocities reaching up to 110 km/s. 

ALMA observation of NGC 5643 was performed on April 19th, 2018 (PID: 2017.1.00082.S, PI: Santiago Garcia-Burillo), targeting the molecular tori to study its characteristics in Seyfert 2 AGN. This project was proposed to map the circumnuclear region using CO(3-2) and HCO$^+$(4-3) of 10 Seyfert galaxies, including NGC 7582, which is also one of the samples in this paper. The telescope pointing was at RA = \(\rm{14^h 32^m 40.778^s}\) and Dec = \(\rm{-44^d 10^m 28.600^s}\) with the continuum sensitivity of 0.0496 mJy/beam. The data calibration was done using CASA version 5.1.1-5.

\subsubsection{NGC 6240}
NGC 6240 is an Ultraluminous Infrared Galaxy (ULIRG) that hosts a double-nucleus AGN, first identified by \citet{Komossa2003-bq} using X-ray spectroscopy with the \textit{Chandra} X-ray Observatory. High-resolution imaging revealed the presence of a double AGN, confirmed by the detection of two prominent neutral Fe $\rm{K\alpha}$ lines. However, recent research has uncovered a third nucleus, indicating a triple AGN system in the final stages of merging \citep{Kollatschny2020-gh}. This finding, confirmed through MUSE observations, resolved the previously identified double nucleus region. Additionally, \citet{Treister2020} reported a bulk of molecular gas within approximately 1" of the region between the two nuclei, as traced by ALMA Band 6 observations of $\rm{^{12}}$CO(2-1) emission.

NGC 6240 was observed on May 4th, 2015 on the Cycle 2 project (PID: 2013.1.00813.S, PI: Naseem Rangwala) with the telescope pointing centred at RA = $\rm{16^h 52^m 58.861^s}$ and Dec = $\rm{+02^h 24^m 03.550^s}$, achieving angular resolution of 0.176 arcseconds and continuum sensitivity of 0.1258 mJy/beam. The reduction was done using CASA version 4.3.1 and calibrated using the pipeline. The observations were conducted under the project to map the warm gas morphologies of the galaxy that undergoing a merging state of luminous infrared galaxies.

\subsubsection{NGC 7582}
NGC 7582 is a nearby Seyfert 2 galaxy exhibiting a Seyfert/ starburst composite nature \citep{Wold2006-ws}. Its nucleus is heavily obscured by dust, likely associated with a torus structure \citep{Regan1999}. X-ray studies, including observations by NuSTAR, indicate that NGC 7582 contains a hidden nucleus obscured by a torus covering approximately 80\%–90\% of the line of sight \citep{Rivers2015-vl}. Additionally, \citet{Braito2017-tv} provide a high-resolution map of the nucleus from \textit{Chandra} observations, showing that the soft X-ray emission originates from a hybrid gas ionised both by intense circumnuclear star formation and the central AGN. Furthermore, NGC 7582 has been extensively studied for its dusty molecular gas content using ALMA, as explored in recent works by \citet{almeida2022diverse} and \citet{Garcia-Burillo2021-ll}.

 Observations towards NGC 7582 were carried out on August 30th, 2018 (PID: 2013.1.00813.S, PI:Santiago Garcia-Burillo). The telescope was pointed at RA = \(\rm{23^h 18^m 23.621^s}\) and Dec = \(\rm{-42^d 22^m 14.060^s}\) with the continuum sensitivity reached 0.0687 mJy/beam. Data processing and analysis were conducted using CASA software, version 5.1.1-5. The project aims to investigate the presence and characteristics of molecular tori in Seyfert galaxies, which are thought to obscure the broad-line region around supermassive black holes in Type-2 AGN.

\subsubsection{NGC 613}
One of the earliest observations on this galaxy was carried out by \citet{Blackman1981} using 17 long-slit spectrograms. Coincidentally, he also observed the galaxies NGC 1097 and NGC 1365. Observations in radio wavelengths on this galaxy have also been conducted using the Very Large Array (VLA) such as in Hummel et al. (1987). More recent work using the ALMA is exemplified by \citet{Miyamoto2017}. In it, the authors studied NGC 613 using the ALMA Bands 3 and 7, which covered the frequency range of 90 – 350 GHz. The molecules observed included CO(3-2), HCN(1-0), and others in the circumnuclear disk and ring in the nucleus of the galaxy. A work by \citet{Audibert2019-lo} suggested a filamentary structure that connects the nuclear to the spiral arms, highlighting a two-arm trailing nuclear spiral, a circumnuclear star-forming ring, and evidence of AGN-driven molecular outflows. Their findings also indicate efficient angular momentum loss, facilitating inflow of molecular gas toward the central black hole.

NGC 613 observations were conducted on November  23rd, 2016, a cycle 3 ALMA observation (PID: 2016.1.00296.S, PI: Francoise Combes). This observation was centred on RA = $\rm{01^h 34^m 18.235^s}$ and Dec = $\rm{-29^d 25^m 06.560^s}$ with the angular resolution of 0.290 arcsec. The reduction and calibration were done using CASA version 4.7.0-1. Under the same project, NGC 613 and NGC 1566 were observed to explore the various conditions in the AGN and star formation environment.

\subsubsection{NGC 1097}
We also examined the barred spiral galaxy NGC 1097, located in the constellation Fornax. This galaxy hosts a LINER/Seyfert 1 nucleus and was initially classified as a LINER (Low-Ionisation Nuclear Emission Line Region) by \citet{heckman1980optical}. A double-peaked Balmer line feature was later discovered by \citet{storchi1993double}. Due to the low bolometric luminosity ($L_{Bol} = 8.6 \times 10^{41} \rm{erg s^{-1}}$, \citet{nemmen2006radiatively}), this galaxy also classified as low-luminosity AGN (LLAGN, bolometric luminosity, $L_{Bol} \lesssim 10^{42} \rm{erg s^{-1}}$, \citet{ho2008nuclear}). \citet{Prieto2019} studied this galaxy across infrared to ultraviolet wavelengths and reported that the observed star formation rate is inconsistent with the available gas reservoir. Using ALMA CO(3–2) observations, \citet{Izumi2017} mapped the cold gas distribution and suggested that the torus in NGC 1097 is thinner compared to that in NGC 1068 \citep{garcia2016alma}. Additionally, dynamical modelling of the CO(3–2) velocity field indicates that the cold molecular gas is concentrated in a thinner layer than the hot gas, which is traced by the 2.12 $\mu$m H\(_2\) emission, within and around the torus.

For NGC 1097, the ALMA observation was carried out on September 14, 2016 (PID: 2015.1.00126.S, PI: Takuma Izumi) on cycle 3 project, reaching angular resolution of around 0.099 arcseconds. The velocity resolution was 3.301 km/s, and the field of view was approximately 16.694 arcseconds. The continuum sensitivity was 0.0304 mJy/beam, and the line sensitivity was 0.748 mJy/beam. The Vlsr used for this observation was 1259.22 km/s. The observation was centred on RA = $\rm{02^h 46^m 19.059^s}$ and Dec = $\rm{-30^d 16^m 29.680^s}$ 

\subsubsection{NGC 1566}
NGC 1566 is a nearby barred spiral galaxy exhibiting Seyfert characteristics \citep{shobbrook1966} and later classified as a Seyfert 1. It is one of the brightest galaxies in the Dorado group, with a centrally symmetric structure as observed in spectroscopic and kinematic studies \citep{aguero2004ngc}. Recent study has revealed a “changing-look” feature in the AGN of NGC 1566, marked by dramatic outbursts across multiple wavelengths and shifts in optical and X-ray properties. Although it is classified as a changing-look AGN, some studies suggest it harbours a low-luminosity AGN (LLAGN) \citep{koribalski2004}. \citet{smajic2015n} investigated the LLAGN the nuclear disk of NGC 1566 using ALMA and SINFONI, revealing significant molecular hydrogen within the disk ($r=3"$) and a spiral structure extending through disk. Also, the work by \citet{combes2014alma} used ALMA CO(3–2) observations to probe the inner structure of NGC 1566, revealing a molecular trailing spiral spanning 50 to 300 pc, located at the inner Lindblad resonance (ILR) ring. This spiral efficiently channels gas toward the nucleus, suggesting AGN fueling through spiral-driven inflows. Their analysis of HCN(4–3) and HCO$^+$(4–3) emission indicates that star formation dominates over AGN heating.

The ALMA observation of NGC 1566 was conducted on November 24, 2016 (PID:2016.1.00296.S , PI: Francoise Combes), a cycle 4 project, and achieved an angular resolution of approximately 0.315 arcseconds. The velocity resolution of the observation was around 0.827 km/s, and the field of view was roughly 16.706 arcseconds. The continuum sensitivity of the observation was about 0.0608 mJy/beam, while the line sensitivity was approximately 1.532 mJy/beam. The Vlsr (local standard of rest velocity) used for this observation was 1502.42 km/s.

\subsubsection{NGC 1808}
NGC 1808 is a nearby Seyfert 2 galaxy located at a distance of 9.5 Mpc \citep{tully2016}. It is identified as a barred galaxy, as revealed by the distribution of neutral hydrogen \citep{phillips1993nuclear}, and is also classified as a starburst galaxy. Notably, it exhibits peculiar "hot spots" in its central 500 pc star-forming region, which contain supernova remnants and young star clusters \citep{sersic1965peculiar, galliano2008embedded}. NGC 1808 has been extensively studied for its gas kinematics and structural features using ALMA observations \citetext{e.g., \citealp{salak2016gas,Audibert2021-aq,chen2023star}}. These studies have confirmed the presence of a compact circumnuclear disk (CND) within \( r < 200 \, \text{pc} \) and a distinct nuclear spiral structure traced by dense gas tracers, including the HCN(4$-$3), HCO$^+$(4$-$3), and CS(7$-$6) lines.

The observation for NGC 1808 was carried out on August 12, 2016 (PID: 2015.1.00404.S, PI: Francoise Combes), which on cycle 3 as Additional Representative Images for Legacy (ARI-L) products. The angular resolution achieved was 0.150 arcseconds with the velocity resolution of 3.298 km/s. The telescope pointing was centred at RA = $\rm{05^h 07^m 42.343^s}$ and Dec = $\rm{-37^d 30^m 46.980^s}$. The data reduction was done using CASA version 4.5.3 while for ARI-L processing, the  CASA version used was 5.6.1-8. The observation of NGC 1808 was part of the project that characterised and quantified the molecular outflows of 20 AGNs on $\sim$100 pc scales.

\subsubsection{NGC 6300}
NGC 6300 is classified as a Seyfert 2 galaxy \citep{phillips1983nearby} and exhibits SBb-type morphology. Located in the southern hemisphere, this ring-barred galaxy features a clumpy, non-uniform torus revealed through X-ray observations from instruments such as Suzaku, the Chandra X-ray Observatory, and the Nuclear Spectroscopic Telescope Array (NuSTAR) \citep{jana2020probing}. These observations suggest that the nuclear region is Compton-thin, with properties such as line-of-sight column density, intrinsic luminosity, and the Fe K$\alpha$ emitting region showing variability and potential evolution over time. A previous study by \citet{awaki2005variability} detected rapid time variability through timing analysis using XMM-Newton data. This variability indicates that NGC 6300 hosts a Seyfert 1 nucleus obscured by dense material surrounding it. These findings further support the unified AGN model, where classification is primarily determined by the degree of obscuration.

The ALMA observation of NGC 6300 was carried out on 22 August 2018 (PID: 2017.1.00082.S, PI: Santiago Garcia-Burillo), achieving an angular resolution of 0.544 arcseconds. The velocity resolution was 3.302 km/s, and the field of view was around 16.684 arcseconds. The continuum sensitivity was 0.0576 mJy/beam, and the line sensitivity was 1.393 mJy/beam. The Vlsr used for this observation was 1100.19 km/s. The observation was centred at RA = $\rm{17^h 16^m 59.473^s}$ and Dec = $\rm{-62^d 49^m 13.980^s}$. The CASA version used for calibration was CASA version 5.1.1. 

\subsubsection{NGC 7314}
NGC 7314, located in the constellation Piscis Austrinus, hosts a nucleus with classifications ranging from Seyfert 1.9 to Seyfert 2. While some studies classify it as Seyfert 1 \citep{stauffer1982nuclear, morris1985ccd, schulz1994long}, others identify it as Seyfert 2 \citep{nagao2000high, trippe2010multi}, reflecting the complexity and variability of its nucleus, as demonstrated by the variability of X-rays (\citealp[e.g.][]{yaqoob1996discovery, turner1997asca}). A recent study of X-ray variability using XMM-Newton observations explored this variability in depth through power spectral density (PSD) analysis, a method used to examine the frequency-dependent behavior of these variations \citep{emmanoulopoulos2016extensive}. The findings suggest that the variability follows a model with a bending frequency, indicating distinct behaviors above and below this frequency. Furthermore, \citet{da2023closer} conducted a detailed multiwavelength analysis of the nuclear region of NGC 7314, investigating properties such as emission-line spectra, the morphology of the line-emitting regions, and spectral variability.

ALMA observation of NGC 7314 was conducted on August 30, 2018 (PID:2017.1.00082.S ,PI: Santiago Garcia-Burillo), with an angular resolution of approximately 0.330 arcseconds. The telescope pointing was centred at RA = $\rm{22^h 35^m 46.230^s} $, Dec = $\rm{-26^d 03^m 00.900^s}$. The velocity resolution was 3.306 km/s, and the field of view was around 16.703 arcseconds. The continuum sensitivity was 0.0682 mJy/beam, and the line sensitivity was 1.616 mJy/beam. The Vlsr used for this observation was 1423.51 km/s. The data calibration was done using CASA version 5.1.1-5 and Pipeline-CASA51-P2-B r40896. Under the same project with NGC 5643, NGC 7582, and NGC 6300, their main goal is to map the CO(3-2) and HCO$^+$ lines within the circumnuclear region of 10 nearby Seyfert galaxies, utilising data from ultra-hard X-ray selected AGN  in the Galaxy Activity, Torus, and Outflow Survey (\citealp[GATOS;][]{Garcia-Burillo2021-ll, GarcaBurillo2024})  sample.

\subsection{Data Reduction}
We utilised pipeline data retrieved from the ALMA Science Archive, focusing on the primary beam-corrected CO(3-2) spectral line data. Once the Flexible Image Transport System (FITS) files were obtained, we applied a 3$\sigma$ clipping to the data cubes to eliminate unwanted background sources, distinguishing them from the central sources of interest. This process also enhanced the signal-to-noise ratio. The final images were much clearer, with reduced background emission. All analyses were conducted using the Astropy SpectralCube package.

Since the ALMA observations for each sample were conducted during different cycles and used different versions of the CASA software, we opted not to re-calibrate the data. Instead, we used the calibrated pipeline data available in the archive, specifically searching for CO(3-2) emission line cube data. The line detection was confirmed using the ALMA Data-Mining Toolkit (ADMIT), accessible through the cube web preview on the ALMA Science Archive. All calibrated pipeline data were corrected for the primary beam.


\section{Results and Discussions}
\label{sec: result}
\subsection{Integrated Intensity Maps}
To analyse the distribution of CO(3-2) around the nucleus, we generated the integrated intensity map (moment 0 map) for all samples using the SpectralCube task \texttt{cube.moment(order=0)}. We used the noise-reduced data cubed to generate the moment 0, producing cleaner and lower noise results. To study the gas distribution around the nucleus, we focused on the nuclear region, covering an area from 0.2 to 6 arcseconds. We then collapsed over the entire range of velocity channels in each data cube to create moment 0 maps over this region. The parameters used are provided in Table \ref{tab:region_velocity}.

\begin{table*}[hbt!]
\caption{Region sizes and velocity widths used in the observed sample.}
\label{tab:region_velocity}
\begin{threeparttable}
    \begin{tabular}{p{3cm}cccc}
    \toprule
    Source Name & \multicolumn{2}{c}{Region Size} & Velocity Width \\
    & (deg) & (arcsec) & (km/s) \\
    \midrule
    \textbf{Compton-Thick} \\
    Circinus Galaxy & 0.001667 $\times$ 0.001667 & 6.000 $\times$ 6.000 & 235.000 - 1283.390 \\
    NGC 1068 & 0.000061 $\times$ 0.000061 & 0.220 $\times$ 0.220 & -400.000 - 280.000 \\
    NGC 4945 & 0.001497 $\times$ 0.001497 & 5.390 $\times$ 5.390 & -257.184 - 1366.301 \\
    NGC 5643 & 0.000989 $\times$ 0.000989 & 3.560 $\times$ 3.560 & 398.826 - 2014.197 \\
    NGC 6240 & 0.001044 $\times$ 0.001044 & 3.760 $\times$ 3.760 & -500.000 - 480.000 \\
    NGC 7582 & 0.001675 $\times$ 0.001675 & 6.030 $\times$ 6.030 & 757.917 - 2373.377 \\
    \midrule
    \textbf{Non Compton-Thick} \\
    NGC 613 & 0.000542 $\times$ 0.000542 & 1.950 $\times$ 1.950 & 1195.216 - 1770.591 \\
    NGC 1097 & 0.000550 $\times$ 0.000550 & 1.980 $\times$ 1.980 & 455.395 - 2069.845 \\
    NGC 1566 & 0.001389 $\times$ 0.001389 & 5.000 $\times$ 5.000 & 704.350 - 2314.350 \\
    NGC 1808 & 0.000467 $\times$ 0.000467 & 1.680 $\times$ 1.680 & 201.742 - 1820.268 \\
    NGC 6300 & 0.000758 $\times$ 0.000758 & 2.730 $\times$ 2.730 & 304.768 - 1920.293 \\
    NGC 7314 & 0.000775 $\times$ 0.000775 & 2.790 $\times$ 2.790 & 621.571 - 2237.008 \\
    \bottomrule
    \end{tabular}
\end{threeparttable}
\end{table*}

\begin{figure*}[htb!]
	\centering
	\begin{subfigure}{1\linewidth}
		\includegraphics[width=0.33\linewidth]{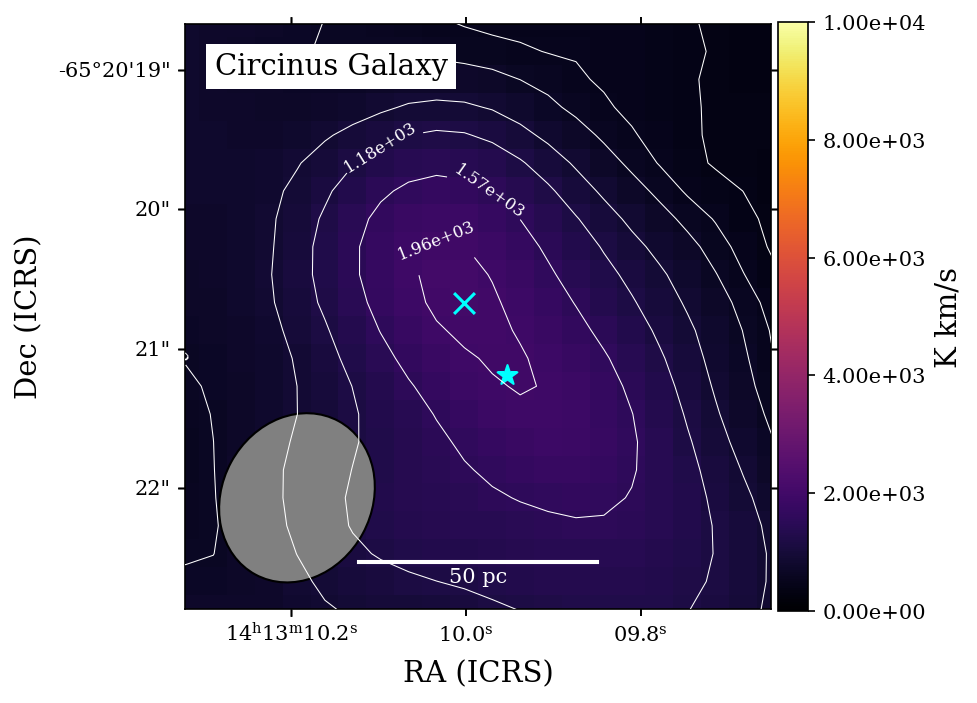} 
        \includegraphics[width=0.33\linewidth]{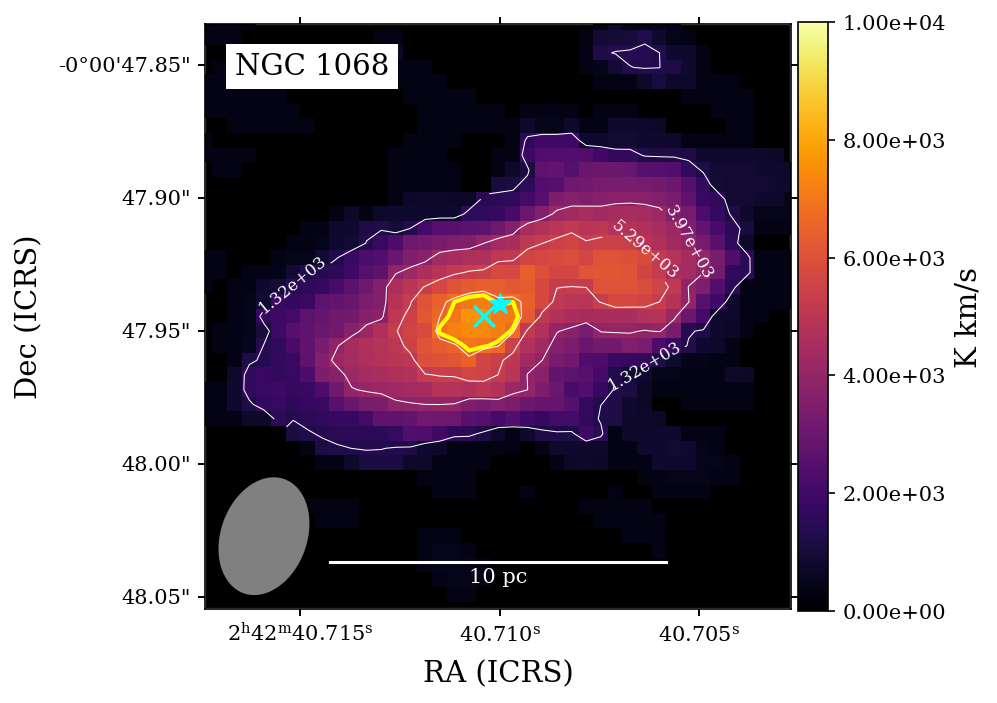} \includegraphics[width=0.33\linewidth]{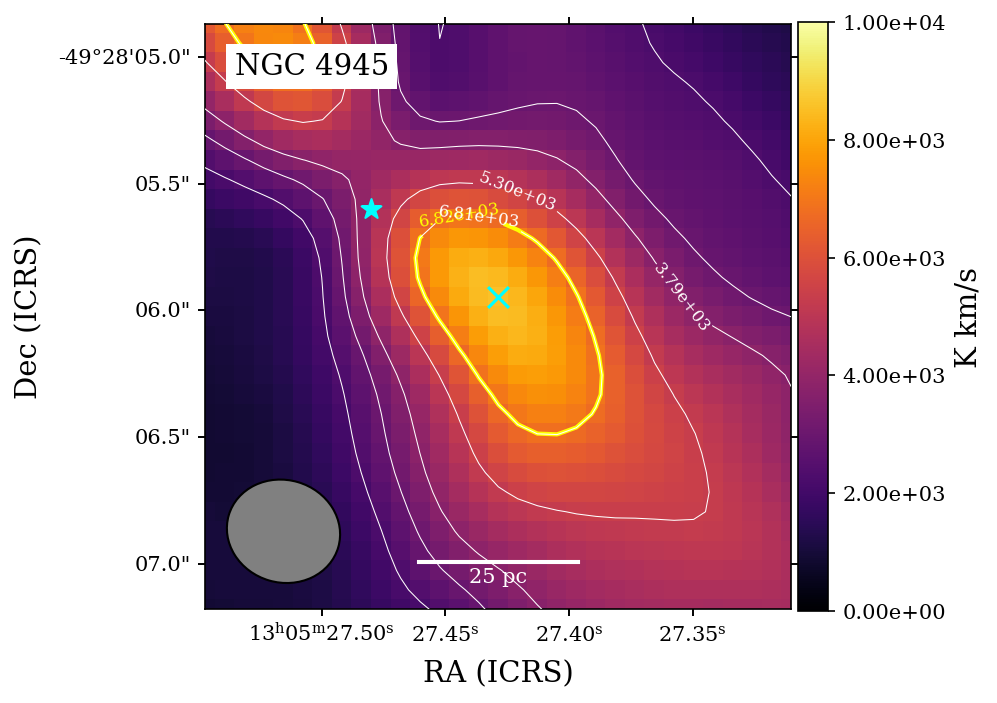} \\
        \includegraphics[width=0.33\linewidth]{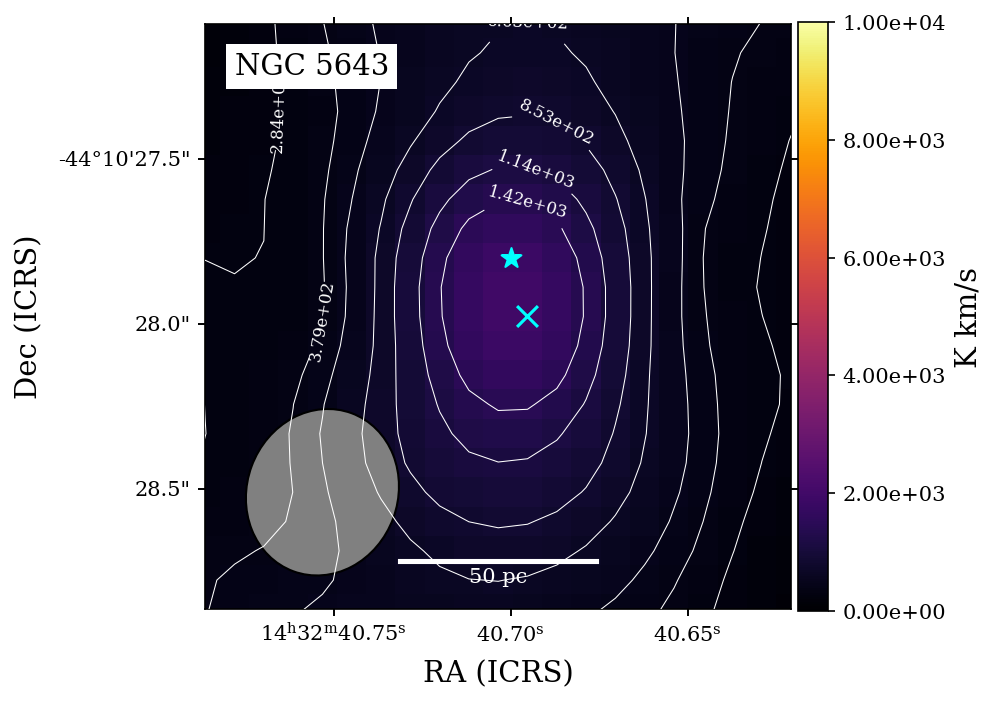} 
        \includegraphics[width=0.33\linewidth]{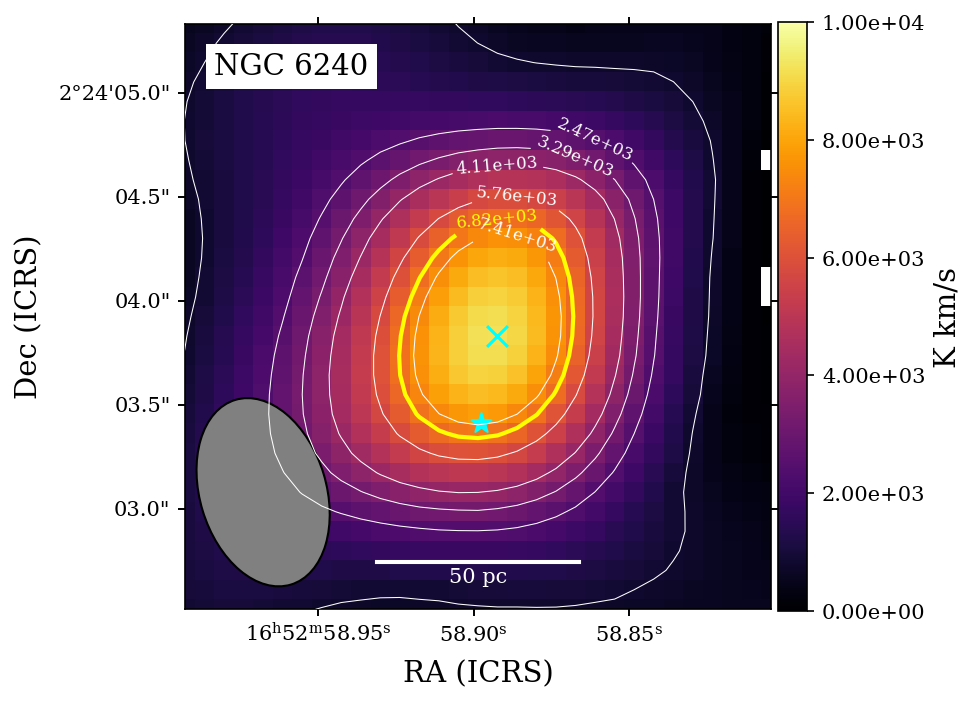} \includegraphics[width=0.33\linewidth]{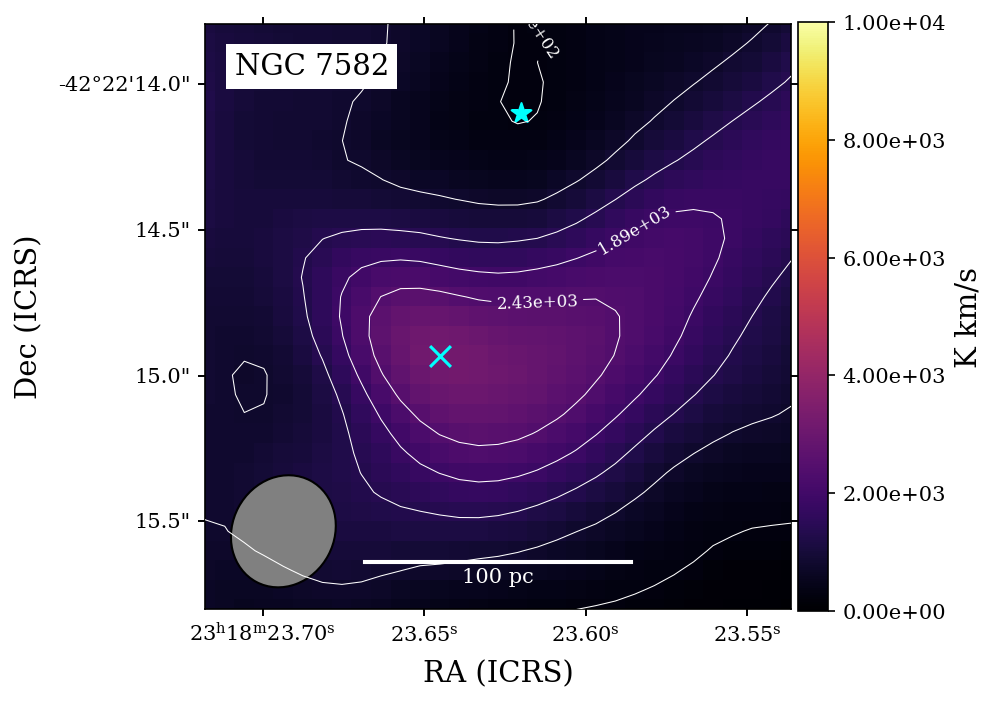}
		\caption{CTAGN sources}
		\label{fig:mom0-ctagn}
	\end{subfigure} \\
	\begin{subfigure}{1\linewidth}
		\includegraphics[width=0.33\linewidth]{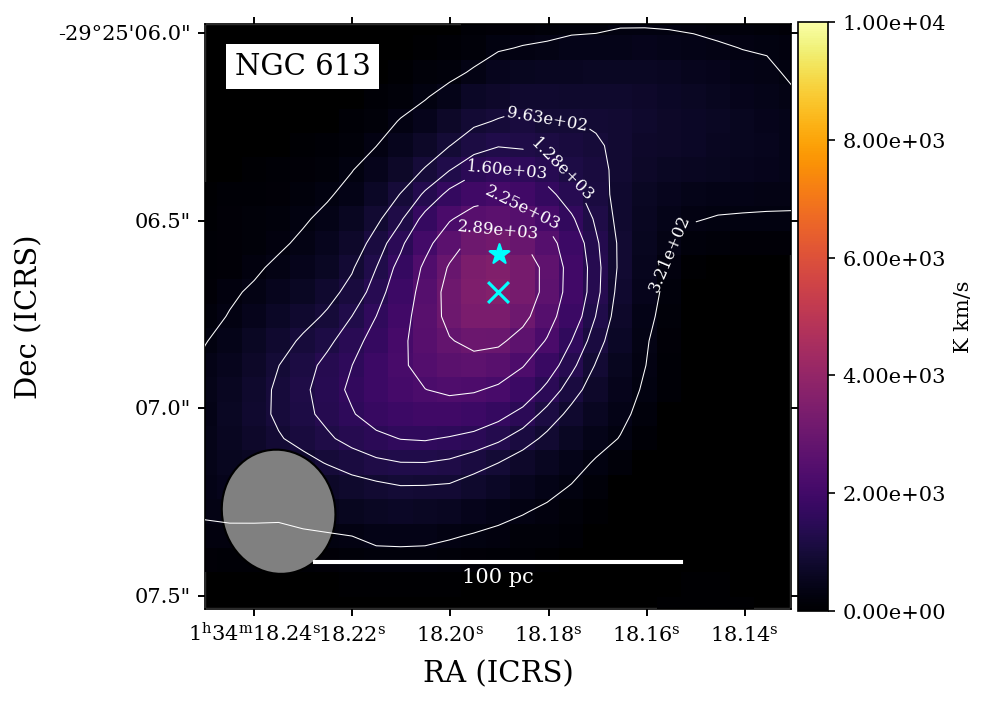}
        \includegraphics[width=0.33\linewidth]{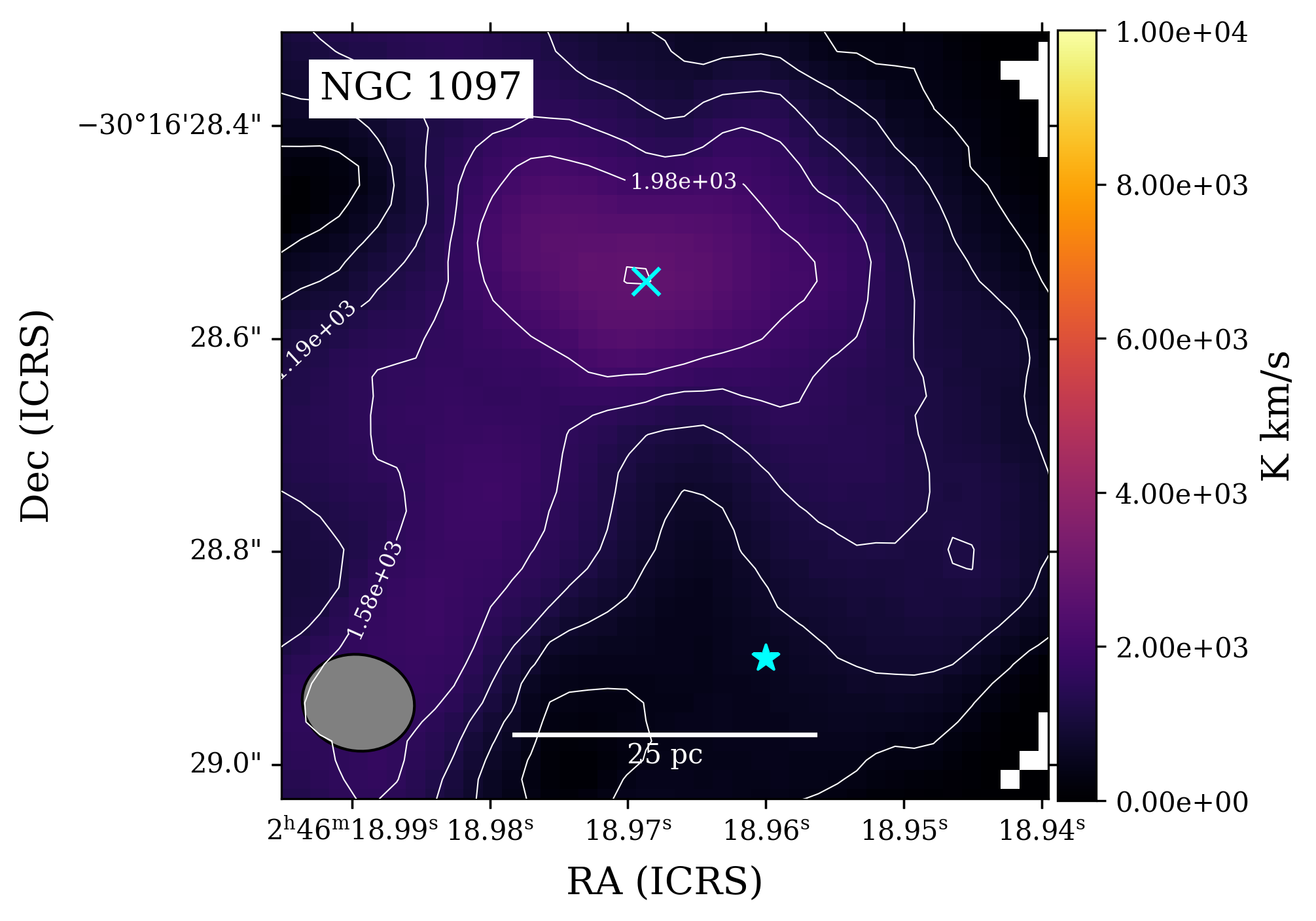} \includegraphics[width=0.33\linewidth]{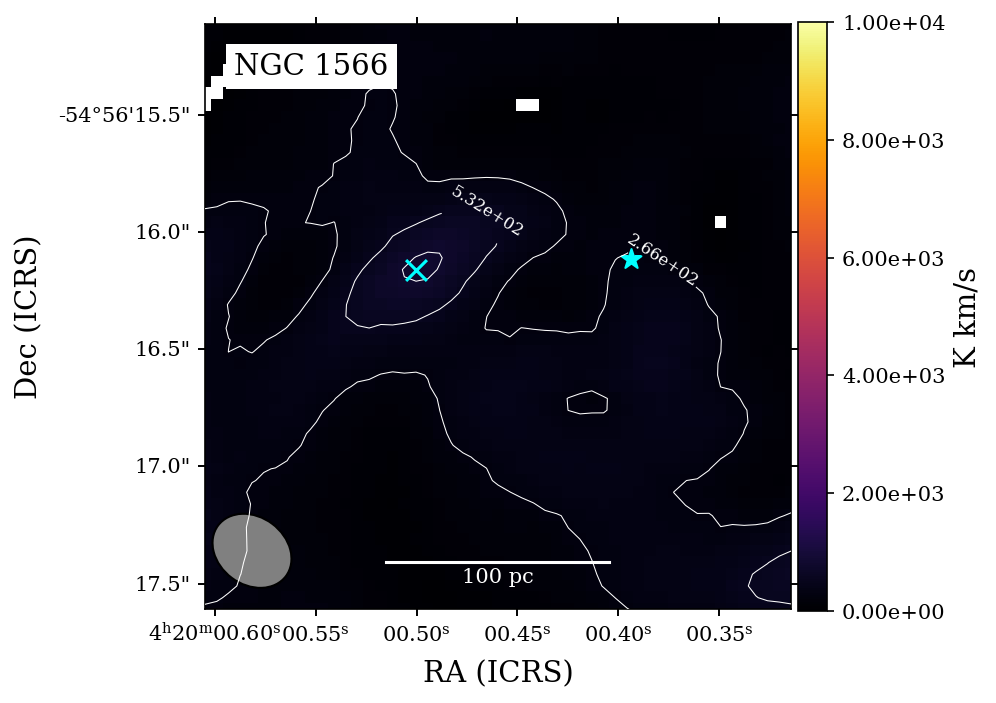} \\
        \includegraphics[width=0.33\linewidth]{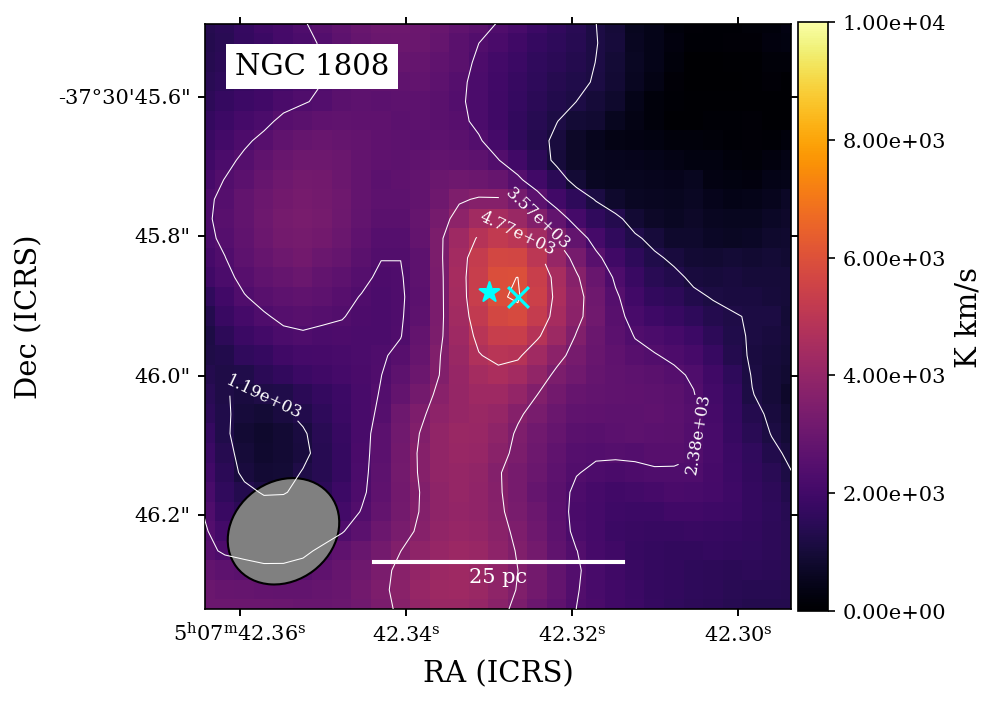}
        \includegraphics[width=0.33\linewidth]{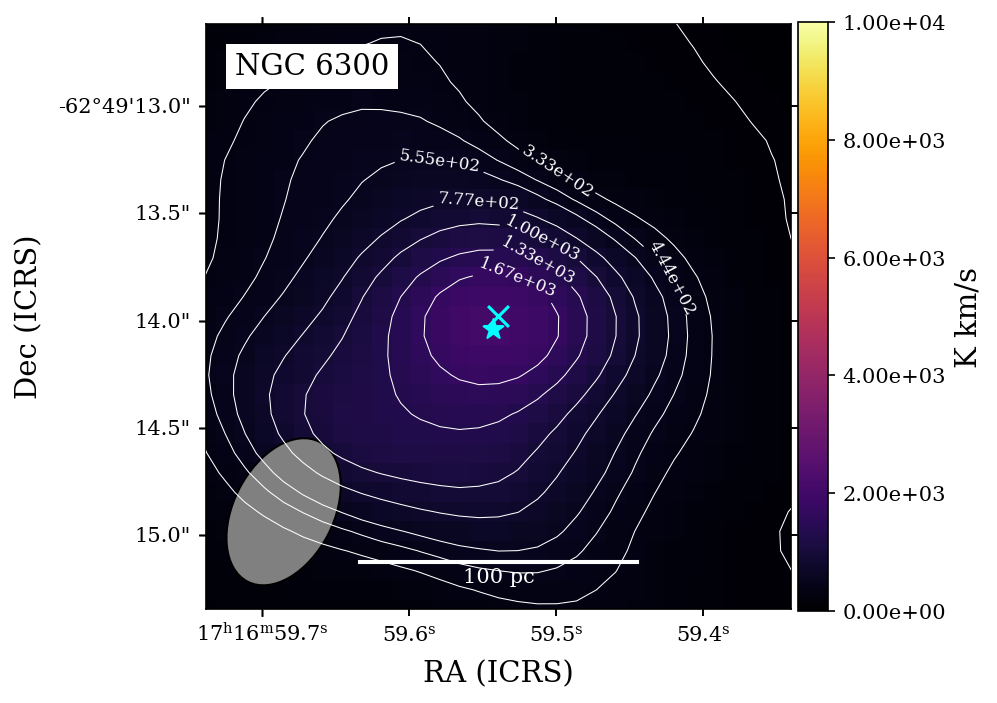} \includegraphics[width=0.33\linewidth]{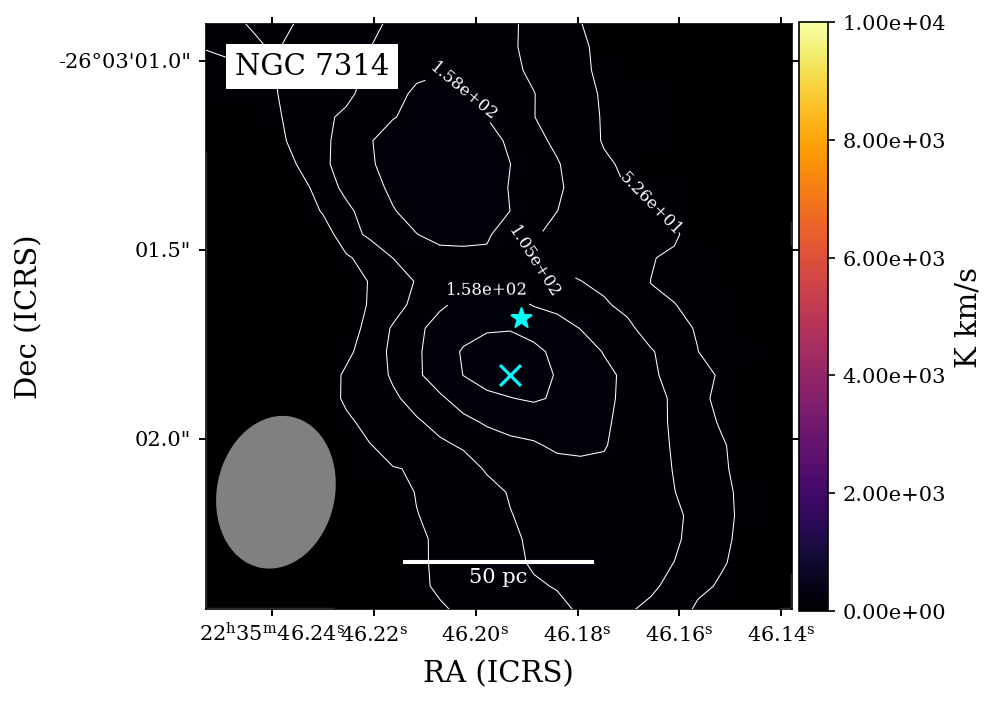} 
        \caption{Non-CTAGN sources}
		\label{fig:mom0-nonctagn}
	\end{subfigure} \\
\caption{CO(3-2) Integrated Intensity (Moment 0) of CTAGN and non-CTAGN sources. The cyan cross marks the integrated CO(3-2) peak emission while the cyan star indicates the AGN position obtained from past literature. The beam size are represented by a grey ellipse. The CTAGN region is shown by a yellow contour. Note that not all sources exhibit a CTAGN yellow contour, as the integrated intensity values in these cases do not meet the threshold typically associated with CTAGN classification.}
    \label{fig:mom0}
\end{figure*}

Figure \ref{fig:mom0-ctagn} and \ref{fig:mom0-nonctagn} present the integrated intensity maps for the CTAGN and non-CTAGN sources. The size of each galaxy region varies according to the extent of the nuclear region; thus, we standardised the region size to a square pixel grid. To ensure a clear and consistent comparison of the CO(3–2) emission distribution and concentration, we standardised the intensity scale values to $I_{\rm{CO_{min}}} = 0\ \rm{K\ km\ s^{-1}}$ and $I_{\rm{CO_{max}}} = 10^4\ \rm{K\ km\ s^{-1}}$. AGN positions, primarily determined through radio observations and referenced from previous literature, are marked by cyan stars, while the emission peaks in the moment 0 maps are indicated by cyan crosses. The yellow contour marks the CTAGN region, where we derived the corresponding threshold for the $I_{\rm{CO}}$ from the column density to identify the CTAGN regions. The $N_{\rm{H_2}}$ values were calculated using a conversion factor, which is discussed in detail in Section \ref{molecular nh2}.

A comparison between the AGN positions from the literature and the CO(3–2) integrated intensity peaks reveals notable spatial offsets. This study aims to explore the spatial relationship between the CO(3–2) intensity peak, which we define as the brightest pixel in the integrated intensity map, and the AGN positions documented in the literature. It is worth noting that the CO(3–2) peak does not necessarily align with the AGN position, as several ALMA studies have shown that AGN positions are often traced by the ~300 GHz continuum peak, while the CO(3–2) emission can be influenced by other processes such as star formation, circumnuclear rings, or nuclear spirals \citetext{e.g. \citealp{izumi2018circumnuclear,Audibert2021-aq,tristram2022alma,combes2023fueling}}. Table \ref{tab:agn_positions_offsets} lists all the AGN positions derived from the literature alongside the separation offsets between these positions and the CO(3–2) integrated intensity peaks.

CO(3–2) is a well-known dense gas tracer commonly used to map the distribution of molecular gas in obscured environments, such as those surrounding AGN \citep[e.g.,][]{Tsai2012, Miyamoto2017, AlonsoHerrero2018, Kawamuro2019-ti, Lamperti2020-ei, jones2023}. It provides valuable insights into AGN-related processes, including radiation and outflows, which can significantly affect the gas reservoir of the host galaxy \citep{Circosta2021-kg}. Additionally, CO(3–2) is particularly useful for tracing cold dense molecular gas, which forms key components of structures such as the torus and circumnuclear disk (CND).

The CO(3–2) moment 0 maps generated for the observed galaxies highlight distinct morphological differences between CTAGN and non-CTAGN sources. For CTAGN sources, such as NGC 1068, NGC 4945, and NGC 6240, the CO(3–2) emission is strongly concentrated in the nuclear regions, with peaks located near the AGN positions. This indicates a significant accumulation of cold molecular gas around the AGN, likely associated with circumnuclear material, such as a torus. Notably, NGC 1068 also exhibits extended structures, including faint, elongated features that could be linked to AGN-driven outflows. In contrast, other CTAGN sources, such as the Circinus galaxy, NGC 5643, and NGC 7582, show weaker emission concentrated near the AGN positions but also display extended components. These features may reflect the presence of nuclear spiral arms or star-forming regions, which could contribute to the observed gas outflows. For non-CTAGN sources, the CO(3–2) emission appears more extended and less concentrated in the nuclear regions compared to CTAGN sources. A relatively low concentration of CO(3–2) is clearly observed in NGC 1566 and NGC 7314, further supporting the idea that the cold molecular gas is distributed across larger areas rather than being confined to the centre. In cases such as NGC 1097 and NGC 1566, the emission exhibits irregular distributions, with peaks offset from the AGN positions, reflecting the dynamic nature of the gas in these systems.

Figure \ref{fig:mom0-ctagn} demonstrates stronger CO(3–2) concentration around the nuclear regions of CTAGN, while Figure \ref{fig:mom0-nonctagn} illustrates the more dispersed emission characteristic of non-CTAGN. These differences align with the understanding that CTAGN are surrounded by dense, thick obscuring material, leading to compact emission in the nuclear region. Although extended structures are present in both CTAGN and non-CTAGN, they appear to be more pronounced in non-CTAGN. This may be due to less dense surrounding material in non-CTAGN, allowing for a broader distribution of the cold molecular gas.

In previous related studies, the Galaxy Activity, Torus, and Outflow Survey (\citealp[GATOS,][]{Garcia-Burillo2021-ll, GarcaBurillo2024}) investigated the dusty molecular tori properties of nearby Seyfert galaxies. The first results of this project, presented by \citet{Garcia-Burillo2021-ll}, revealed spatially resolved CO(3-2) and HCO$^+$(4-3) emissions in disks surrounding AGN. While CO(3-2) is frequently detected in AGN, its emission typically does not peak at the centre. Instead, it is more commonly observed in regions surrounding the AGN, often associated with structures such as rings or the edges of cavities formed by outflows. This could explain the notable offset of CO(3-2) peak integrated intensity, $I_{\rm{CO}}$ from the AGN position obtained from the previous literature, shown in Figure \ref{fig:mom0}. Furthermore, the study quantified the central concentration of molecular hydrogen (H$_2$) on various spatial scales, linking H$_2$ surface density with AGN luminosities and Eddington ratios. A deficit of molecular gas was identified in the nuclear regions, particularly in AGN with higher luminosities and Eddington ratios. This finding suggests that AGN feedback, which is stronger in more luminous AGN, plays a crucial role in redistributing or depleting molecular gas from the nuclear regions.

\begin{table*}[htb]
    \centering
    \caption{AGN Positions and Offsets}
    \label{tab:agn_positions_offsets}
    \begin{threeparttable}
        \begin{tabular}{p{7em}ccccccc}
        \toprule
        Source Name & \multicolumn{2}{c}{AGN Position\tnote{1}} & \multicolumn{2}{c}{Peak Intensity\tnote{2}} & $I_{\rm{CO}}$ & Separation Offset & Ref.\tnote{3} \\
        & RA (h:m:s) & Dec (d:m:s) & RA (h:m:s) & Dec (d:m:s) & ($\times 10^3\ \rm{K\ kms^{-1}}$) & (arcsec) & \\
        \midrule
        \multicolumn{8}{l}{\textbf{Compton-Thick AGN}} \\
        Circinus Galaxy & 14:13:09.948 & -65:20:21.187 & 14:13:10.002 & -65:20:20.670 & 2.0482 & 0.6173 & 1 \\
        NGC 1068 & 02:42:40.711 & -00:00:47.940 & 02:42:40.710 & -00:00:47.944 & 7.6091 & 0.0101 & 2 \\
        NGC 4945 & 13:05:27.477 & -49:28:05.570 & 13:05:27.429 & -49:28:05.947 & 8.5583 & 0.6039 & 3 \\
        NGC 5643 & 14:32:40.700 & -44:10:27.800 & 14:32:40.704 & -44:10:27.888 & 1.9713 & 0.1842 & 4 \\
        NGC 6240 & 16:52:58.896 & +02:24:03.395 & 16:52:58.892 & +02:24:03.832 & 9.2432 & 0.4278 & 5 \\
        NGC 7582 & 23:18:23.600 & -42:22:13.000 & 23:18:23.645 & -42:22:14.931 & 3.2204 & 0.8766 & 6 \\
        \midrule
        \multicolumn{8}{l}{\textbf{Non-Compton-Thick AGN}} \\
        NGC 613 & 01:34:18.19 & -29:25:06.59 & 01:34:18.190 & -29:25:06.690 &  3.6042 & 0.1000 & 7 \\
        NGC 1097 & 02:46:18.96 & -30:16:28.900 & 02:46:18.973 & -30:16:28.536 & 2.7749 & 0.3912 & 8 \\
        NGC 1566 & 04:20:00.397 & -54:56:16.625 & 04:20:00.384 & -54:56:16.560 & 0.8314 & 0.1269 & 9 \\
        NGC 1808 & 05:07:42.33 & -37:30:45.88 & 05:07:42.327 & -37:30:45.887 & 6.0154 & 0.0420 & 10 \\
        NGC 6300 & 17:16:59.473 & -62:49:13.38 & 17:16:59.539 & -62:49:13.980 & 2.0782 & 0.0649 & 11 \\
        NGC 7314 & 22:35:46.23 & -26:03:00.90 & 22:35:46.193 & -26:03:01.892 & 0.2371 & 0.2141 & 12 \\
        \bottomrule
        \end{tabular}
        \tnote{1} {Adopted AGN positions from past literature.} \\
        \tnote{2} {Integrated CO(3-2) emission peak position.} \\
        \tnote{3} {Reference on the AGN position. (1) \cite{Greenhill2003-gt}, (2) \cite{Garcia-Burillo2019-qo}, (3) \cite{Greenhill_1997}, (4) \cite{Greenhill2003}, (5) \cite{Hagiwara2011-hc}, (6) \cite{Andonie2022-bv}, (7) \cite{Audibert2019-lo}, (8) \cite{Hummel1987}, (9) \cite{Ricci2017}, (10) \cite{Audibert2019-lo}, (11) \cite{Garcia-Burillo2021-ll}, and (12) \cite{Evans2010-nd}}
    \end{threeparttable}
\end{table*}

\subsection{Molecular Hydrogen Column Density}
\label{molecular nh2}
The  moment 0 peak values were determined, and their corresponding positions are provided in Table 4. The moment 0 map of CO(3–2) highlights a concentrated emission near the nucleus, suggesting the presence of a highly obscuring medium within the size of approximately 10 to 100 pc around the nucleus. Using the moment 0 CO(3-2) peak integrated intensity value, We derived molecular hydrogen column density, $N_{\mathrm{H_2}}$ using CO-to-$\mathrm{H_2}$ conversion factor. We converted the intensity value from Jy/beam to $\rm{K}$ using the standard brightness temperature relation on the original cube file before proceed with moment 0 maps. Equation \eqref{eq:1} shows the method of deriving the molecular hydrogen column density, $N_{\rm{H_2}}$ obtained from \citet{Rahmani2016}.

\begin{equation}
\label{eq:1}
N_\mathrm{H_2}(cm^{-2})=X_{\mathrm{CO}} \times I_\mathrm{CO}
\end{equation}
where $N_{\mathrm{H_2}}$ is the molecular hydrogen column density, $X_\mathrm{CO}$ is the CO-to-$\mathrm{H_2}$ conversion factor, and $I_\mathrm{CO}$ is the CO integrated intensity in K kms$^{-1}$. We adopted a constant Milky Way conversion factor, $X_{\mathrm{CO}} = 2.2 \times 10^{20} \mathrm{cm}^{-2} \ (\mathrm{K\ km\ s}^{-1})^{-1}$ from \citet{Solomon1983} which corresponds to $\alpha_{\mathrm{CO}}=3.5 \ M_\odot \ \mathrm{pc}^{-2} \ (\mathrm{K\ km\ s}^{-1})^{-1}$.

\begin{figure}[!ht]
    \centering
    \includegraphics[width=1\linewidth]{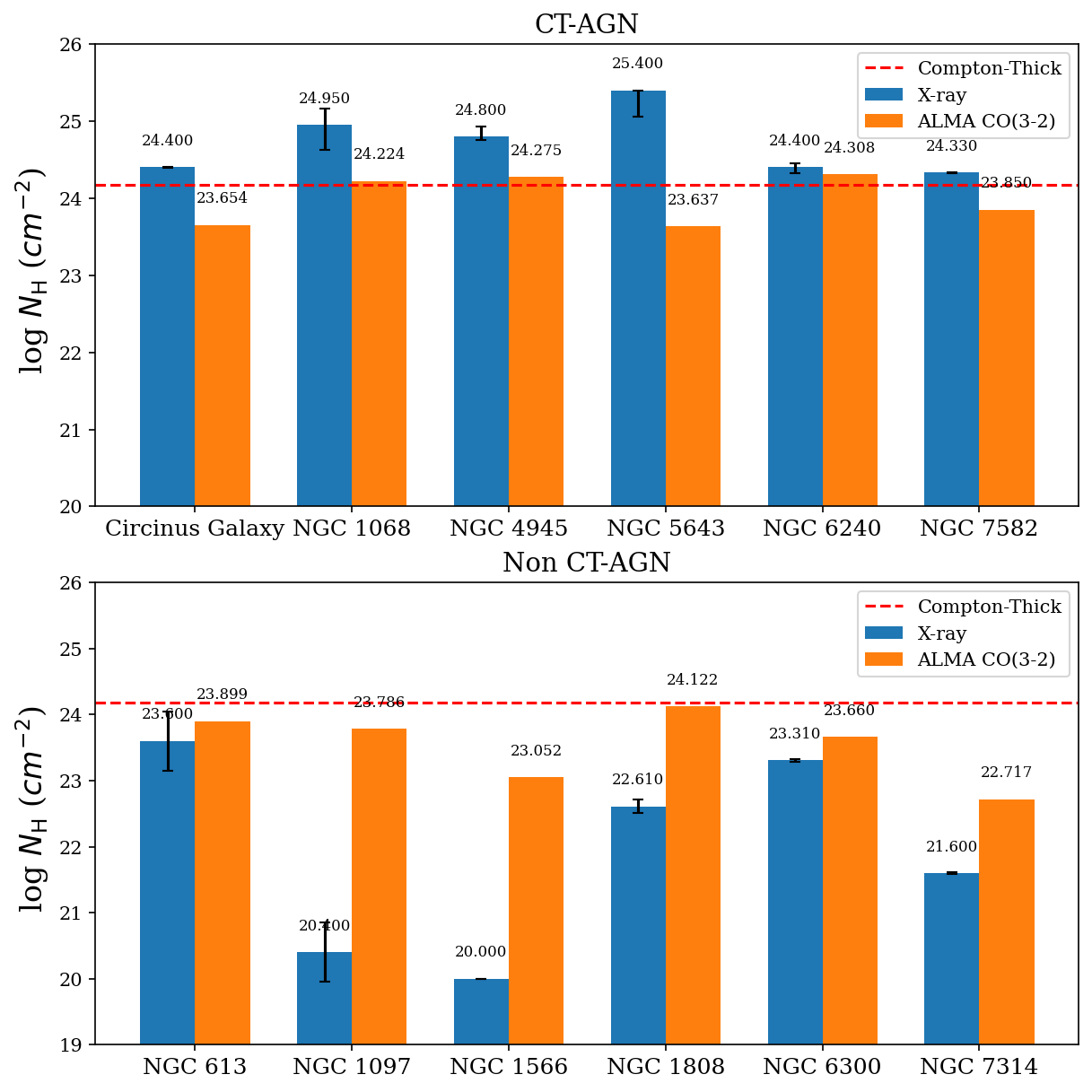}
    \caption{Column densities for molecular hydrogen, $N_{\rm{H_2}}$ derived using CO-to-$\rm{H_2}$ conversion factor and total hydrogen, $N_{\rm{Htotal}}$ as inferred from X-ray. The red dotted line marks the CTAGN threshold.}
    \label{fig:nh}
\end{figure}

\begin{table}[h!]
    \centering
    \caption{Molecular hydrogen column densities, $N_{\rm{H_2}}$ and logarithmic values for CTAGN and non-CTAGN. The $N_{\rm{H_2}}$ are given in units of $10^{23} \, \mathrm{cm}^{-2}$. The data highlights the relationship between sub-mm and X-ray observations, offering insights into the physical conditions within these galaxies.}
    \label{table:nh2}
    \begin{threeparttable}
    \begin{tabular}{l c c c}
        \toprule
        Source Name & $N_{\rm{H_2}}$ & ${\rm{log_{10}}} \ N_{\rm_{H_2}}$  &${\rm{log_{10}}} \ N_{\rm_{H}}$ \\ 
        & ($\times \ \rm{10^{23} \ cm^{-2}}$) & ($\rm{cm^{-2}}$)  & ($\rm{cm^{-2}}$)\\
        \midrule
        Circinus Galaxy & 4.51 & 23.65 & 24.40\tnote{1}\\
        NGC 1068 & 16.74 & 24.22 & 24.95\tnote{1}\\ 
        NGC 4945 & 18.83 & 24.27 & 24.80\tnote{1} \\ 
        NGC 5643 & 4.34 & 23.64 & 25.40\tnote{1}\\ 
        NGC 6240 & 20.34 & 24.31 & 24.40\tnote{1}\\ 
        NGC 7582 & 7.09 & 23.85 & 24.33\tnote{1}\\ 
        NGC 613 & 7.93 & 23.90 & 23.60\tnote{3}\\ 
        NGC 1097 & 6.10 & 23.79 & 20.40\tnote{5}\\ 
        NGC 1566 & 1.13 & 23.05 & 20.00\tnote{2}\\ 
        NGC 1808 & 13.23 & 24.12 & 22.61\tnote{4}\\ 
        NGC 6300 & 4.57 & 23.66 & 23.31\tnote{2}\\ 
        NGC 7314 & 0.52 & 22.72 & 21.60\tnote{2}\\ 
        \bottomrule
    \end{tabular}   
    \tnote{1}{Taken from \citet{Ricci2015}} \\
    \tnote{2}{Taken from \citet{Ricci2017}} \\ 
    \tnote{3}{Taken from \citet{she2017}}\\
    \tnote{4}{Taken from \citet{Heike_2007}} \\
    \tnote{5}{Taken from \citet{Asmus2015}} 
    \end{threeparttable}
\end{table}

The derived molecular hydrogen column density, $N_{\rm{H_2}}$ values for the sources are listed in Table \ref{table:nh2}. We compared the derived molecular hydrogen column densities, $N_{\rm{H_2}}$ with the X-ray-derived column densities, $N_{\rm{h}}$ from the 70-Month \textit{Swift}/BAT catalogue, except for NGC 613, NGC 1097, and NGC 1808. For these sources, we used $N_{\rm{H}}$ values from previous literature: NGC 613 from \citet{she2017} as observed by \textit{Chandra}, NGC 1097 from \citet{Asmus2015}, and NGC 1808 from \citet{Heike_2007}. 

In Figure \ref{fig:mom0}, the CO moment 0 maps illustrate variations in the morphology and concentration of dense CO gas near the nuclear region. From the $I_{\rm{CO}}$ we can infer the $N_{\rm{H_2}}$ values and identify the CTAGN regions. The yellow contours outline the boundaries of the CTAGN regions, visible in NGC 1068, NGC 4945, and NGC 6240. The high concentration of CO gas within these nuclear regions results in high $N_{\rm{H_2}}$ values that exceed the CTAGN threshold. This allows us to define the extent of the CTAGN region, characterised by substantial obscuration from dust and gas. In contrast, for other cases, the absence of yellow contours indicates that the $I_{\rm{CO}}$ values do not surpass the CTAGN threshold, reflecting lower concentrations of molecular gas.

The results, shown in Figure \ref{fig:nh}, indicate that the molecular hydrogen column densities, $N_{\rm{H_2}}$ for CTAGNs are generally lower than the X-ray-derived column densities, $N_{\rm{H}}$, consistent with previous studies. This discrepancy suggests that $N_{\rm{H}}$ may overestimate the total amount of obscuring material. The difference arises because X-ray and sub-mm observations trace different components and absorbers. X-ray emission is typically associated with inverse-Compton scattering originating in the corona, while sub-mm emission such as CO may arise from colder, absorbing gas. The lower $N_{\rm{H_2}}$ values in CT-AGN could also result from the thick optical depth of the surrounding material, as CO emission, commonly used to infer $\rm{H_2}$ densities, becomes optically thick, leading to underestimations of the $N_{\rm{H_2}}$ compared to X-ray-derived values, $N_{\rm{H}}$ that account for all type of hydrogen (atomic, HI and molecular, $\rm{H_2}$) along the line of sight.

In contrast, non-CTAGN exhibit higher molecular hydrogen column densities, $N_{\rm{H_2}}$, compared to the total hydrogen column densities, $N_{\rm{H}}$. This can be attributed to their less dense surrounding environments, allowing for more effective tracing of $\rm{H_2}$ through CO emission. Additionally, isotopologues of CO, such as $\rm{^{13}CO}$, are generally optically thin in less dense molecular clouds \citep{tan201112co, shimajiri2014high, barnes2020lego}. In the case of CTAGN, the surrounding environment consists of highly dense regions, leading to high column densities. Under such conditions, CO emission can become saturated, limiting its effectiveness in tracing molecular gas. This phenomenon is demonstrated by \citet{Shetty2011-jv}, where they conducted radiative transfer modelling using magnetohydrodynamics (MHD) simulations of molecular clouds (MCs). Their study suggested that beyond a certain column density threshold, the CO line emission no longer correlates with the molecular mass. This limitation also extends to the estimation of $H_2$ column densities, as it relies on the same $X_{\rm{CO}}$ conversion factor. This supports the idea that the CO line may be optically thin in non-CTAGN regions.

The other possible factor for the discrepancy between $N_{\rm{H_2}}$ and $N_{\rm{H}}$ could be due to the total hydrogen column density, $N_{\rm{H}}$ derived from X-ray observations includes contributions from both atomic hydrogen (HI) and molecular hydrogen ($\rm{H_2}$). In environments with significant amounts of HI, which is not traced by CO, the total hydrogen column density, $N_{\rm{HTotal}}$, can exceed the molecular hydrogen column density, $N_{\rm{H_2}}$, derived from CO observations. This can be explained by the positive correlation between $N_{\rm{HI}}$ and $N_{\rm{H}}$ observed by \citet{Ostorero2010-oq, Ostorero2017-iu}, where higher hydrogen column densities, $N_{\rm{H}}$, measured from X-rays correspond to more significant HI detection. In this context, we can conclude that high $N_{\rm{H}}$ in CTAGN corresponds to higher HI detection, while lower $N_{\rm{H}}$ in non-CTAGN results in reduced HI detection. Thus, it will explain the discrepancy observed in this work.

In Fig.~\ref{fig:nh}, most of the \(N_{\rm{H_2}}\) does not exceed the threshold typically used to identify Compton-thick AGN (CTAGN), which is primarily defined through X-ray observations ($N_{\mathrm{H}} \geq 1.5 \times 10^{24} \ \rm{cm^{-2}}$). This threshold represents the presence of highly dense regions in the nucleus, where the obscuring material around the AGN reaches extreme column densities. If \(N_{\rm{H_2}}\) were to be used alone for identifying CTAGN, it may require a revised threshold. Only NGC 1068, NGC 4945, and NGC 6240 show \(N_{\rm{H_2}}\) values that exceed this limit, indicating such dense regions in their nuclei. For both CTAGN and non-CTAGN, the difference between \(N_{\rm{H_2}}\) and the threshold is relatively small, with values staying above \(N_{\rm{H}} = 10^{22} \ \mathrm{cm^{-2}}\). On the other hand, the total hydrogen column density ,\(N_{\rm{H}}\) shows a larger difference from the CTAGN threshold, especially for non-CTAGN sources.

In comparison with previous studies, the initial results of the GATOS project \citep{Garcia-Burillo2021-ll} also explored the variability of molecular gas masses within the nuclear region. They derived molecular hydrogen column densities, $N_{\rm{H_2}}$, for several galaxies, some of which overlap with this study (e.g., NGC 1068, NGC 5643, NGC 6300, NGC 7314, and NGC 7582). The $N_{\rm{H_2}}$ values reported in their work are slightly lower than those presented here. This discrepancy could be attributed to the different CO-to-H$2$ conversion factors, $X_{\mathrm{CO}}$ used in the analyses. While the GATOS project adopted a Milky Way conversion factor of $X_{\mathrm{CO}} = 2 \times 10^{20} \ \mathrm{cm}^{-2} \ (\mathrm{K \ km \ s}^{-1})^{-1}$, this study uses a slightly higher $X_{\mathrm{CO}}$, resulting in marginally higher $N_{\rm{H_2}}$ values.

\subsection{Correlation between \texorpdfstring{$N_{\rm{H_2}}$ and $N_{\rm{H}}$}{NH2 and NH}}

\begin{figure}[ht]
    \centering
    \includegraphics[width=1\linewidth]{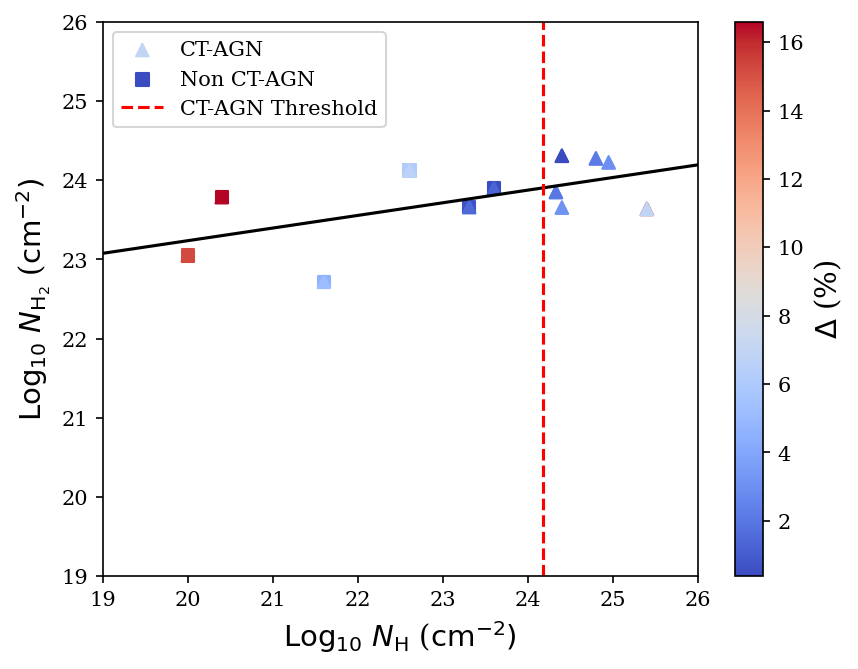}
    \caption{Correlation between the derived molecular hydrogen column density, $N_{\rm{H_2}}$ and total hydrogen column density, $N_{\rm{Htotal}}$. The CTAGNs are marked with triangular shape and square for non-CTAGNs. The colour bar on the right shows the percentage difference of $N_{\rm{H_2}}$ from $N_{\rm{Htotal}}$. The linear regression fit is represented by the solid black line.}
    \label{fig:nh2-corr}
\end{figure}

In a previous study from \citet{Ostorero2017-iu}, they carried out a correlation analysis between radio column densities ($N_{\rm{HI}}$) and X-ray column densities ($N_{\rm{H}}$), suggesting a significant positive correlation, which implies as increasing value of X-ray observation, there will be increasingly detection in HI observation. In this work, to investigate the relationship between derived molecular hydrogen column density, $N_{\rm{H_2}}$ and total hydrogen column density, $N_{\rm{H}}$, we performed a correlation analysis using three different correlation test, which is Kendall's correlation analysis, Spearman's correlation test, and Pearson's correlation analysis. Figure \ref{fig:nh2-corr} displays the column densities plot between $N_{\rm{H}}$ and $N_{\rm{H_2}}$ on a log scale. with the linear regression fit line.

First, we investigate the correlation by applying Kendall's correlation test to the galaxy sample. The test shows that there is a weak positive monotonic association between  $N_{\rm{H}}$ and $N_{\rm{H_2}}$ with Kendall's tau value, $\tau = 0.26$. This indicates that there is a slight tendency of higher  $N_{\rm{H}}$ value to be associated with high  $N_{\rm{H_2}}$ value. However, the correlation is not strong. Given the $P = 0.24$ is above 0.05, which common significance level, the relationship is not statistically significant. Thus, we are unable to claim there is a meaningful monotonic relationship between  $N_{\rm{H}}$ and $N_{\rm{H_2}}$. Further investigation with a larger dataset might be necessary to measure the significant relationship between these two column density. 

To further study the correlation, we applied Spearman's correlation to the data sets, showing a moderate positive monotonic association between both column densities. The correlation value, $r_s = 0.42 $ with $P = 0.17$ suggest that even with a positive moderate correlation, it is not strong enough to be considered reliable based on this sample alone. Larger data sets might be needed to confirm any potential relationship between these column densities. The correlation results shown by both tests agree with each other.  Both tests show a positive monotonic correlation whereas a stronger correlation is shown by Spearman's test. The difference however slightly stronger than Kendall's test, which is expected because Spearman's test is more sensitive to monotonic relationships. Both p-values are greater than the common significance level of 0.05, indicating that neither test found a statistically significant correlation. However, the p-value from Spearman's test is closer to the significance threshold than that from Kendall's test, suggesting a marginally stronger but still statistically non-significant trend. This correlation has also been studied by \citet{Garcia-Burillo2021-ll}, who reported a statistically significant positive relationship between these two variables, confirmed through Pearson's correlation test. Other correlation tests such as Kendall and Spearman also suggest a similar result. The observed positive trend, though not statistically significant, suggests a possible association between dense gas (traced by CO(3-2)) and higher X-ray column densities. This could indicate the presence of molecular gas in regions with high obscuration, but further studies are required to confirm this relationship.

We investigate the relationship of these column densities by applying a regression analysis. We fit the data using linear regression fit, showing the linear relation given by;
\begin{equation}
\label{eq:nh_reg}
{\rm{log_{10}}} \ N_\mathrm{H_2}(\rm{cm^{-2}})= a \ {\rm{log_{10}}} \ N_{\mathrm{H}} + b
\end{equation}
as shown in figure \ref{fig:nh2-corr}. Although the correlation test shows a weak positive correlation, we performed the regression analysis to observe the dependency of  $N_{\rm{H_2}}$ towards $N_{\rm{H}}$. We obtained the slope, $a = 0.16$ with the intercept value, $b = 20.04$, showing alignment with the correlation, with the r-squared value expected to be low as weak or moderate correlation is shown. The R-squared value of 0.35 indicates a moderate fit, which could explain that the fit only captures some of the variability of the data while a significant amount of the remains unexplained. This preliminary analysis could potentially explore future works that can be done on the sample, to provide a better relationship of the column densities that are the main parameter of identifying CTAGN. 

\section{Conclusion}
\label{sec: conclusion}

We present ALMA observations of CO(3-2) towards the central region of AGN samples, covering the Compton-thick and non-Compton-thick sources. We consider nearby AGN samples based on their classification (CTAGN or non-CTAGN) and redshift, $z \leq 0.005$. We select the CTAGN samples from \citet{Ricci2017}, in the 70-Month \textit{SWIFT}/BAT catalogue, as well as the non-CTAGN sources. We also obtain samples from \citet{she2017} for both CTAGN and non-CTAGN. We search for ALMA CO(3-2) observations on the ALMA science archive and utilize the pipeline data provided, which results in having 6 CTAGNs and 6 non-CTAGNs. To obtain a good signal-to-noise ratio, we applied 3$\sigma$ clipping before performing an integrated intensity map (moment 0). By using the Milky Way CO-to-$\rm{H_2}$ conversion factor, we derive the molecular hydrogen column density ($N_{\rm{H_2}}$), by taking the CO(3-2) peak integrated intensity. In order to study the correlation between molecular hydrogen column density ($N_{\rm{H_2}}$) and total hydrogen column density ($N_{\rm{H}}$) inferred from X-ray observation, we perform correlation test using Kendall's and Spearman's test. In addition, we fit a linear regression to explore the relationship between both column densities. Our main results are summarised as follows:
\begin{enumerate}
    \item Dense Region Tracing: The dense regions traced by CO(3-2) exhibit emission peaks near the nucleus, which align with AGN locations identified in previous studies, with minimal offsets. This suggests the presence of concentrated dense regions around the nucleus, that could be associated with dusty molecular torus.
    \item Comparison of $N_{\rm{H_2}}$ and $N_{\rm{H}}$: In CTAGN samples, $N_{\rm{H_2}}$ tends to be lower than $N_{\rm{H}}$. This discrepancy may arise from the different components producing the emissions, and possibly an underestimation of $\rm{H_2}$ due to CO being optically thick in dense regions. Conversely, in non-CTAGN samples, $N_{\rm{H_2}}$ often exceeds $N_{\rm{H}}$. This could be attributed to CO being optically thin in less dense regions, leading to higher estimates of $N_{\rm{H_2}}$.
    \item Correlation Analysis: Both Kendall's (0.2595) and Spearman's (0.4203) correlation tests indicate a positive monotonic relationship between $N_{\rm{H_2}}$ and $N_{\rm{H}}$ which agree with previous studies. This suggests that as $N_{\rm{H}}$ increases, $N_{\rm{H_2}}$ also tends to increase, with Kendall's test showing a weaker association and Spearman's test indicating a moderate positive correlation.
    \item Statistical Significance: The p-values from both tests (Kendall's: 0.2326, Spearman's: 0.1737) indicate that the observed correlations are not statistically significant at the conventional threshold of 0.05. Therefore, the sample does not provide strong enough evidence to confirm a significant correlation between these column densities.
\end{enumerate}
Future studies could focus on refining the CO-to-$\mathrm{H_2}$ conversion factor by incorporating more precise measurements of the CO integrated intensity, $I_{\rm{CO}}$ and the molecular hydrogen column density, $N_{\rm{H_2}}$. Additionally, exploring other methods for determining  $N_{\rm{H_2}}$, such as using other molecular lines or dust emission, could provide further insights into the obscuration properties of CTAGN.

Our findings contribute to a broader understanding of AGN environments and the complexities of measuring obscuration properties. Accurately determining column densities is crucial for distinguishing between different types of AGNs and understanding the nature of their central regions. Further exploration in this area, with more comprehensive datasets and refined methodologies, is essential for advancing our knowledge of the diverse and dynamic processes at play in these powerful cosmic phenomena.

\begin{acknowledgement}
This paper makes use of the following ALMA data: ADS/JAO.ALMA  2013.1.00813.S, 2015.1.00126.S., 2015.1.01286.S, 2015.1.00404.S, 2016.1.00232.S, 2016.1.00296.S, 2017.1.00082.S, and 2018.1.01236.S. ALMA is a partnership of ESO (representing its member states), NSF (USA) and NINS (Japan), together with NRC (Canada) and NSC, ASIAA (Taiwan) and KASI (Republic of Korea), in cooperation with the Republic of Chile. The Joint ALMA Observatory is operated by ESO, AUI/NRAO and NAOJ.

This research was supported by the Faculty of Science Research Grant (GPF081-2020) and the Universiti Malaya Special Grant (RUIP002-2023). We would like to extend our gratitude to both funding bodies for their generous support, which has been invaluable in advancing our research.

\end{acknowledgement}

\section*{Data Availability}
The data underlying this article are publicly available in the ALMA Science Archive at \href{linkALMA}{https://almascience.nrao.edu/aq/}, and can be accessed with the following project IDs: 2013.1.00813.S, 2015.1.00126.S., 2015.1.01286.S, 2015.1.00404.S, 2016.1.00232.S, 2016.1.00296.S, 2017.1.00082.S, and 2018.1.01236.S.


\bibliography{references}

\end{document}